\documentclass[conference]{IEEEtran}

\usepackage{bigstrut}

\usepackage{cite}

\ifCLASSINFOpdf
   \usepackage[pdftex]{graphicx}
   \DeclareGraphicsExtensions{.pdf}
\else
\fi
\usepackage[cmex10]{amsmath}
\usepackage{amssymb}
\usepackage{array}

\usepackage{mdwmath}
\usepackage{mdwtab}

\usepackage{eqparbox}

\usepackage[tight,footnotesize]{subfigure}
\usepackage{graphicx}

\usepackage{fancyhdr}
\begin{document}
%
\IEEEspecialpapernotice{Submitted to the $9^{th}$ USA/Europe ATM R\&D Seminar - Berlin 2011}{}


\title{ATC Taskload Inherent to the Geometry of Stochastic 4-D Trajectory Flows with Flight Technical Errors}

\author{\IEEEauthorblockN{Vlad Popescu, John-Paul B. Clarke, Karen M. Feigh, Eric Feron}
\IEEEauthorblockA{School of Aerospace Engineering\\
Georgia Institute of Technology\\
Atlanta, GA 30332-0150, USA\\
Email: \{vlad.popescu, johnpaul, karen.feigh, feron\} @gatech.edu \\
}
}


%


\maketitle

\begin{abstract}
A method to quantify the probabilistic controller taskload inherent to maintaining aircraft adherence to 4-D trajectories within flow corridors is presented. An Ornstein-Uhlenbeck model of the aircraft motion and a Poisson model of the flow scheduling are introduced along with reasonable numerical values of the model parameters. Analytic expressions are derived for the taskload probability density functions for basic functional elements of the flow structure. Monte Carlo simulations are performed for these basic functional elements and the controller taskload probabilities are exhibited.
\end{abstract}


%
\IEEEpeerreviewmaketitle

\section{Introduction}

Gridlock is forecast to occur both in the U.S and Europe if the current air traffic practices are not changed. It is widely believed that a technological shift is necessary to accommodate this growth. Along with this technical evolution, a change in procedures and infrastructure will take place, that will give more autonomy to aircraft and rely on higher-precision, less fault-tolerant operations and trajectories.

The NextGen concept of operations \cite{JPDO:conops07} envisions \emph{flow corridors}, where aircraft follow high-precision 4-D-metered trajectories, across relatively compact regions of airspace. These Required Navigation Performance (RNP) trajectories with increasingly more stringent precision requirements use equipment that is certified with probabilistic tolerance bounds \cite{PBN08}. Airspace precision standards in the U.S. differ widely depending on the phases of flight and airspace blocks, ranging from an accuracy of 0.1 NM (RNP-0.1) for precision approach-to-landing to 10 NM (RNP-10) in areas of the Pacific Ocean \cite{Radio08}.

Flow corridors rely on self-separation using Automatic Dependent Surveillance-Broadcast (ADS-B). This equipment allows the construction of high-density, high-altitude en route sectors \cite{Mun07}, \cite{You10}. The role of controllers is to monitor traffic and intervene in case of extreme - and potentially dangerous - deviations from planned 4-D trajectories. 

A unified and provable theory linking traffic complexity, airspace block capacity and controller intervention currently does not exist. This research aims to contribute to the unification of these disparate concepts by introducing tools to clarify the dependency between air traffic flow corridor geometry and the controller taskload needed to maintain the structure. \emph{Taskload} is defined here as the rate at which the controller must intervene to return aircraft that have reached the tolerance bounds outside of their 4-D trajectories. After an overview of the state of the art, in Section \ref{sec:model}, we introduce fundamental stochastic models of the aircraft flight (an Ornstein-Uhlenbeck process) and of the flow scheduling (a Poisson process), along with reasonable numerical values of the model parameters.

Precisely estimating the parameters of the models is highly relevant, since overestimating the variability may cause inefficiency while underestimating it may cause unsafe conditions and elevated controller taskload. The Ornstein-Uhlenbeck aircraft model is calibrated to match simulated data, obtained from a random number generator defined by a Johnson unbounded system $S_U$. Using simulated data is required in the absence of experimental navigation recordings; the random number generator provides fictitious data consistent with statistical studies of the aircraft motion. The Poisson flow model uses intensity parameters given in recently published research. 

In Section \ref{sec:analytic} we then derive analytic expressions for the taskload probability density function for basic functional elements of the flow structure: a single lane, multiple parallel lanes, and flow corridor crossings. In Section \ref{sec:sim}, we show taskload probabilities found by Monte Carlo simulations of these basic functional elements.

\section{Background}

Probabilistic models of air traffic are not new to the literature. This is especially true in the case of conflicts and collisions, where the ICAO has in fact defined the acceptable levels of fatal accident risk at one mid-air collision (physical incrossing) per $10^9$ flight hours \cite{ICAO98}. The classic model used to predict risk values for some basic types of air traffic management, known as the Reich collision model, was developed in 1964 \cite{Rei64}.

Over the last few years, Bakker and Blom generalized the Reich model and developed collision risk models based on first hitting times of Markov processes, using hybrid-state Markov processes with switching coefficients, or Petri nets \cite{Blo06}. Irvine showed that the minimum relative displacement between conflicting aircraft can follow a normal distribution, and that the minimum distance between them has a folded normal distribution \cite{Irv01}.

Much of the conflict detection research invokes position uncertainty. The classic characterization of position uncertainty was suggested by Paielli and Erzberger and models the error as normally distributed with an time-wise linear standard deviation of the along-track error and a constant standard deviation of the cross-track error \cite{PaielliErzberger:1997}. This model has been expanded by Blin et al. to include a dynamic resulting from superposed Wiener processes (Brownian motion) affecting position and velocity, or by switching from a position-based to a time-based random process: the time of arrival at a trajectory waypoint becomes the random variable while the position is deterministic \cite{Blin}.

Stochastic conflict detection research is usually geared at designing resolution algorithms. Significant work in this field has been undertaken by Hu, Prandini et al., who have modeled aircraft oscillations around their prescribed trajectories by means of a Brownian motion, the relative position between two aircraft by time-inhomogeneous Markov chains, introduced spatially correlated wind perturbations, and proposed decentralized algorithms that use potential field methodology or the reachability analysis of a switching diffusion model approximated by a Markov chain \cite{Pra08}. 

Despite all the aforementioned research, not much has been done in terms of stochastic traffic flow management and route structure design. Moreau et al. \cite{moreau2005}, or Wan and Roy \cite{Wan09}, compared management strategies using Poisson flow models. The Poisson flow model has nevertheless raised some doubts: Schmidt studied conflicts at route intersections and showed this to be expressible as a sum of correlated random variables with a variance larger than that obtained by the Poisson flow model \cite{Sch77}. Sala\"{u}n et al. have also shown that the circadian variations of the aircraft arrival process do not correspond to a homogeneous Poisson process, but that the arrivals can nevertheless be modeled as a non-homogeneous process over that time frame \cite{Sal10}.

Finally, little research exists concerning the stochastics of controller intervention: Dunlay and Horonjeff showed that subjective conflict risk (perceived by controllers) approximately adheres to a Poisson distribution \cite{Dun75}; Jeddi et al. modeled the influence of air traffic controllers on the aircraft separation (a random variable) by a Gaussian noise, and proposed statistical standards (i.e. bounds on separation, target value, variance) to improve throughput in approach sequences \cite{Jed08}. Although no current theory can unify complexity, capacity, and taskload, Vela et al. have proposed a parametric upper bound on the minimal resolution rate required at a flow intersection \cite{Vel10}.

\section{Model}
\label{sec:model}

\subsection{Aircraft}
We model the aircraft flight motion as an Ornstein-Uhlenbeck mean reverting process oscillating around a deterministic trajectory. The Ornstein-Uhlenbeck process is a stationary Gaussian process with bounded variance and is governed by the stochastic differential equation (\ref{eqn:OU}).

\begin{align}
dX_t=\kappa(\mu-X_t)dt+\sigma dW_t 
\label{eqn:OU}
\end{align}

Here $\mu$ is the mean vector, $\kappa$ the elasticity matrix, $\sigma$ the volatility matrix, and $W_t$ a Wiener process (standard Brownian motion). This model is more complex than the Brownian model found elsewhere in the literature for aircraft uncertainty, and is meant to reproduce the behavior of an imperfect flight management system (modeled in the volatility of the Brownian oscillation), and a controlling cockpit (pilot and guidance equipment). The cockpit attempts to prevent major deviations (modeled in the elasticity) from the deterministic trajectory (the mean). To our knowledge, this model has not been previously used in this context. 

The controller taskload due to managing individual aircraft can be calculated as the number of interventions needed to prevent an aircraft from trespassing the allowed precision bounds. The authors recognize that this is only one aspect of the controller taskload and of the subsequent workload, and also relies on the assumption that the pilot would not become aware of the deviation and to react (hence changing the aircraft dynamics). The probability density function for the first hitting time of a boundary $\tau^{(k)}_1 =\inf\{t:X_t\ge k\}$ for an origin at $X_0$ and a level $k$ by an unidimensional Ornstein-Uhlenbeck process has the closed-form solution (\ref{eqn:hit_density}) given by Leblanc et al. \cite{Leb00}

\begin{align}
&\mathcal P[\tau^{(k)} \in dt]=f_{\tau}(t)= \dfrac{k-\frac{X_0}{\sigma^2}}{\sqrt{2\pi}}( \frac{\kappa}{\sigma^2 \sinh\kappa t} )^{\frac{3}{2}}\times\ldots \notag\\
&\times \exp \left[ \frac{\kappa}{2\sigma^2} \left( (\frac{X_0}{\sigma^2} -\mu )^2- (k-\mu)^2 +\sigma^2 t - (\frac{X_0}{\sigma^2})^2 \coth\kappa t  \right)  \right] \label{eqn:hit_density} 
\end{align}

From this, a measure of the taskload required for controlling an individual aircraft is obtained  as the probability of the number of interventions over a given period $T_{max}$. Assuming all realizations of the aircraft trajectory (after each corrective intervention from the controller) are independent, the (discrete) probability (\ref{eqn:task_density}) is obtained from autoconvoluting the hitting times density:
\begin{align} 
\mathcal P[N^{(T)}\ge n]&=\mathcal P[\tau_1+...+\tau_n \le T_{max}]\notag\\
\cdots &= \int_0^{T_{max}}\left\{\stackrel{n-1}{\star}\right\}f_{\tau}(t)dt \label{eqn:task_density}\end{align} 

Here $f_{\tau}(t)$ is the hitting time density function, $N^{(T)}$ is the number of corrections over time $T_{max}$, $\{\tau_1\cdots\tau_n\}$ are the successive hitting times, and $\left\{\stackrel{n-1}{\star}\right\}$ represents $n-1$ (continuous) autoconvolutions of the density function. 
 \begin{align*}
\left\{\stackrel{0}{\star}\right\}f_{\tau}(t)&=f_{\tau}(t) \\
\left\{\stackrel{1}{\star}\right\}f_{\tau}(t)&=\int_0^t f_{\tau}(x)\cdot f_{\tau}(t-x)dx \\
\left\{\stackrel{k+1}{\star}\right\}f_{\tau}(t)&=\int_0^t \left\{\stackrel{k}{\star}\right\}f_{\tau}(x)\cdot f_{\tau}(t-x)dx
\end{align*}

Therefore, for $n\ge1$ the probability of the number of interventions is (\ref{eqn:ProbFtask}).

\begin{align}
\mathcal P[N^{(T)}= n]&=\int_0^{T_{max}}\left\{\stackrel{n-1}{\star}\right\}f_{\tau}(t)-\left\{\stackrel{n}{\star}\right\}f_{\tau}(t)dt\label{eqn:ProbFtask}\\
\mathcal P[N^{(T)}= 0]&=1-\mathcal P[N^{(T)}\ge 1]\notag \\
\cdots &=1-\int_0^{T_{max}}f_{\tau}(t)dt \label{ProbOtask_ac}
\end{align}

\begin{figure}[!h]
\centering
\includegraphics[width=1.6in]{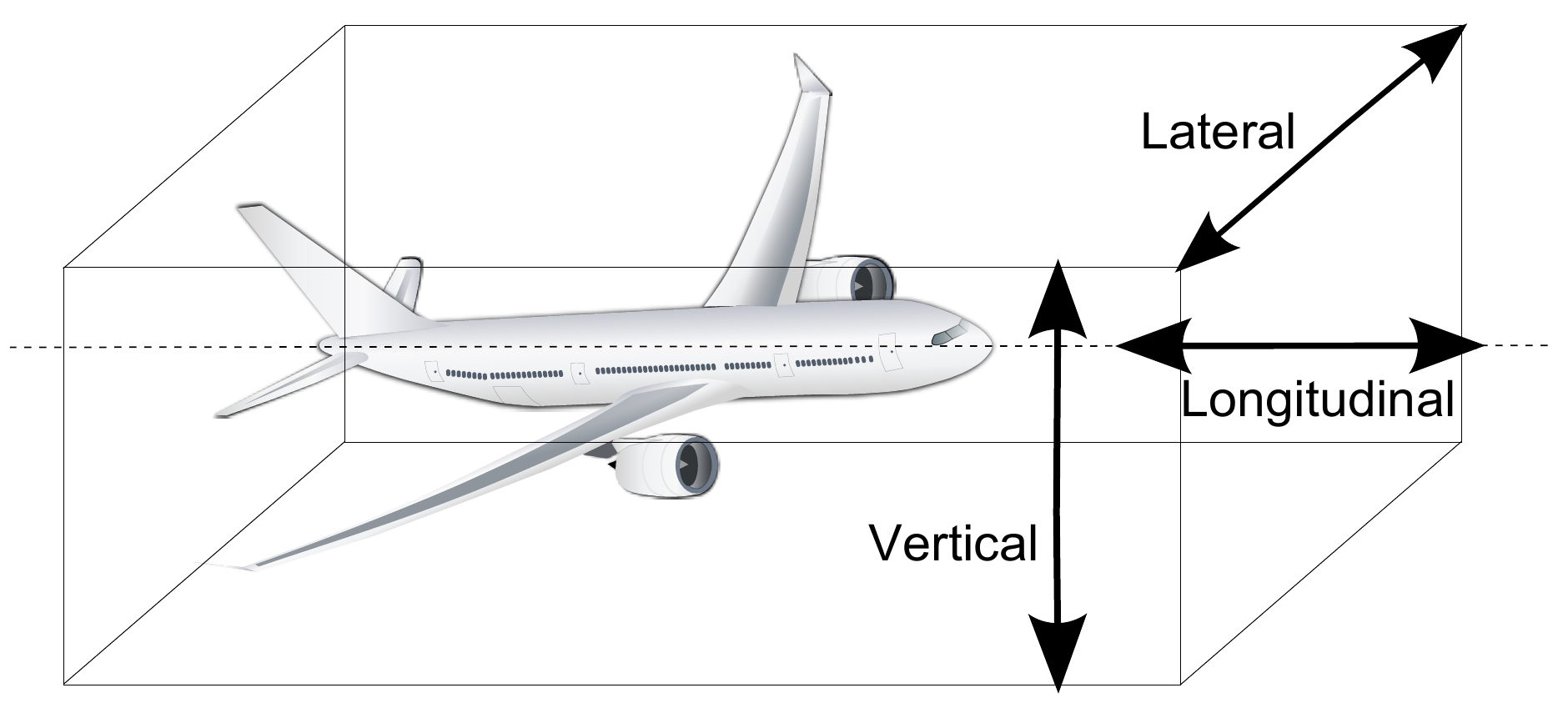}
\caption{Aircraft motion dimensions}
\label{fig:flow_dim}
\end{figure}

But these explicit formulations are only true for an unidimensional stochastic process. In the multidimensional case, no closed-form solution is known for the first hitting time density function of a correlated Brownian motion with drift \cite{Met10}.  Such an n-dimensional stochastic process $X_t$ is a solution to (\ref{eqn:BM}). But through a change of variables and by using the scalability property of the Wiener process, the Ornstein-Uhlenbeck process given in (\ref{eqn:OU}) can be re-written as (\ref{eqn:OUrewrite}).
\begin{align}
dX_t&=\mu dt+\sigma dW_t\label{eqn:BM}\\
dX_{\frac{t}{\sigma^2}}&=-\frac{\kappa}{\sigma^2} X_{\frac{t}{\sigma^2}}dt+dW_t\label{eqn:OUrewrite}
\end{align}
From comparing (\ref{eqn:OUrewrite}) to (\ref{eqn:BM}), for which no closed-form solution of the first hitting time density is known, it becomes apparent that attempting to express hitting time density for a multidimensional correlated Ornstein-Uhlenbeck process is a daunting task. Therefore, we shall use an uncorrelated three-dimensional aircraft model, distinguishing longitudinal along-track motion from lateral and vertical motion (see Figure \ref{fig:flow_dim}). By assuming that the along-track and cross-track deviations are decoupled (an assumption previously used by Paielli and Erzberger \cite{PaielliErzberger:1997}), then implicitly the subjacent unidimesional Ornstein-Uhlenbeck processes can be considered to be uncorrelated, and thus the expressions in (\ref{eqn:hit_density}) and (\ref{eqn:task_density}) can be used. The coupling between vertical and lateral errors will however be implicit to data used for the calibration of the model. Nevertheless, there will be no \emph{a priori} explicit dynamic correlation. Empirically, statistically uncorrelated vertical and lateral error data can be obtained from sampling the two dimensional data at different rates for each dimension.

Some literature exists which can help quantify the various probability parameters.

The ICAO Performance-based Navigation (PBN) manual \cite{PBN08} defines Path Definition Error (PDE), Flight Technical Error (FTE), and Navigation System Error (NSE), for which it assumes independent, zero-mean Gaussian distributions. PDE occurs when the path defined in the RNAV system does not correspond to the path expected to be flown over the ground, and can be considered to be negligible; FTE relates to the air crew or autopilot's ability to follow the defined path or track; NSE refers to the difference between the aircraft's estimated position and actual position. Further ICAO assumptions are that FTE is an ergodic stochastic process within a given flight control mode but that nothing can be said of the NSE due to sensor errors, relative position from navaids, and inertial errors.

Current ICAO standards for PBN demand that the TSE remain equal to or less than the required navigation accuracy (the RNP level) with $95\%$ probability (i.e. $95\%$ of the flight time), and that the TSE has a probability of less than $10^{-5}$ in exceeding twice the required navigation accuracy. Typically, the $10^{-5}$ requirement provides a greater constraint: for a normally distributed cross-track TSE, this bounds the standard deviation to be $\sigma \le 0.45\cdot$RNP, while the $95\%$ requirement sets a bound at $\sigma \le 0.51\cdot$RNP. In addition, onboard equipement is designed to issue an alert if there is a greater than $10^{-7}$ probability per hour that the error exceed the RNP level by two times.

Numerical and statistical parameters for FTE and FMS performance can be found in the literature, mostly dealing with small aircraft and general aviation \cite{Hun93}, \cite{Wil05}. But perhaps the most conclusive statistical analysis was conducted by Levy et al. \cite{Lev03}, who identified a Johnson $S_U$ (\emph{``unbounded system''} \cite{Jon49}) curve as the best fit probability density function for the lateral FTE of a Boeing 747, under RNP-0.3 sensitivity. The fitted curve in (\ref{eqn:JohnsonSU}) is a transformation of the standard normal distribution, and is adapted to heavy-tailed skewed data. An estimate of the curve parameters was conducted by Levy et al. according to a dedicated algorithm developped by Hill et al. in 1976 \cite{Hil76} which matches the first four moments. For the lateral FTE data, the first four moments found by Levy et al. (mean, variance, relative skewness, relative kurtosis) are shown in Table \ref{tab:FTEfit}. The vertical mean and variance are also given by Levy et al., and since no performance data exists for longitudinal data, these were postulated for the purpose of this research by scaling the distribution. If $X_{SNV}$ is a standard normal variate then its transformed $X_{SU}$ obeys (\ref{eqn:JohnsonSU}) and its density function is (\ref{eqn:JSUdensity}).

\begin{align}
X_{SU}&=g(X_{SNV})=\lambda\sinh\left(\frac{X_{SNV}-\gamma}{\delta}\right)+\xi \label{eqn:JohnsonSU}\\
f_{X_{SU}}(x)&=\left|\frac{1}{g'(g^{-1}(x))}\right|f_{X_{SNV}}(g^{-1}(x))\label{eqn:JSUdensity}\\
f_{X_{SNV}}(x)&=\frac{1}{\sqrt{2\pi}}e^{\frac{-x^2}{2}} \notag\\
g'(x)&=\frac{\lambda}{\delta}\cosh(\frac{x-\gamma}{\delta}) \notag \\
g^{-1}(x)&=\delta\sinh^{-1}(\frac{x-\xi}{\lambda})+\gamma\notag
\end{align}

Two other reasonable fitting possibilities are the three-parameter gamma density function (\emph{Pearson Type III}) and the normal distribution. Nevertheless, due to the heavy-tailed data, assuming a normal distribution of the FTE significantly underestimates the risk of transgressing a containment area: according to Levy et al., the ICAO regulatory $10^{-5}$ probability extreme value boundary for a normal distribution is a hundred times more likely to be hit under a Johnson $S_U$ distribution (i.e. with a $10^{-3}$ probability).

\begin{table}
\caption{FTE statistical  moments}
\label{tab:FTEfit}
\centering
\begin{tabular}{|l|c|c|c|}
\hline\hline
Moment & Lateral & Vertical & Longitudinal \bigstrut\\ \hline\hline
$\mu_1$& -0.028 NM & 8 ft & -0.1 NM  \bigstrut \\ \hline
$\mu_2=\sigma^2$ & $9\cdot10^{-4}$ NM$^2$ & 26 ft$^2$ & $2.25\cdot 10^{-2}$ NM$^2$ \bigstrut \\ \hline
$\beta_1=\displaystyle\frac{{\mu_3}^2}{{\mu_2}^2}$ & \multicolumn{3}{c|}{0.243} \bigstrut \\ \hline
$\beta_2=\displaystyle\frac{\mu_4}{{\mu_2}^2}$ & \multicolumn{3}{c|}{5.107} \bigstrut\\ \hline\hline
\end{tabular}
\end{table}

The distribution parameters $\gamma$, $\delta$, $\lambda$, $\xi$ can be approximated from the moments by numerically solving a $24^{th}$ order algebraic equation \cite{Win78}. Without going into further detail, this method provides the parameter values in Table \ref{tab:FTEparam}. The transformed standard normal quartile values shown in Table \ref{tab:quant} can then be deduced from the resulting cumulative density function.

\begin{table}
\caption{Johnson $S_U$ parameters for FTE fit}
\label{tab:FTEparam}
\centering
\begin{tabular} {|c|c|c|c|}
\hline\hline
&Lateral&Vertical&Longitudinal \bigstrut\\ \hline\hline
$\gamma$ & \multicolumn{3}{c|}{0.4566} \bigstrut \\ \hline
$\delta$& \multicolumn{3}{c|}{1.897} \bigstrut \\ \hline
$\lambda$&0.0443 NM& 7.2907 ft& 0.2145 NM \bigstrut \\ \hline
$\xi$&-0.01567 NM& 10.0362 ft& -0.0401 NM \bigstrut \\ \hline\hline
\end{tabular}
\end{table}

\begin{table}
\caption{Johnson $S_U$ quartiles}
\label{tab:quant}
\centering
\begin{tabular} {|c|c|c|c|}
\hline\hline
& Lateral (NM) & Vertical (ft) & Longitudinal (NM)\bigstrut\\ \hline\hline
$Q_1$ & $-6.98\cdot10^{-2}$ & 1.147 & -0.302 \bigstrut \\ \hline
$Q_2$ & $-3.89\cdot10^{-2}$ & 6.215 & -0.152 \bigstrut \\ \hline
$Q_3$ & $-1.46\cdot10^{-2}$ & 10.2 & $-3.52\cdot10^{-2}$ \bigstrut \\ \hline
$Q_4$ & $9.98\cdot10^{-3}$ & 14.27 & $8.42\cdot10^{-2}$ \bigstrut \\ \hline\hline
\end{tabular}
\end{table}

\begin{figure}[!h]
\centering
\includegraphics[width=1.6in]{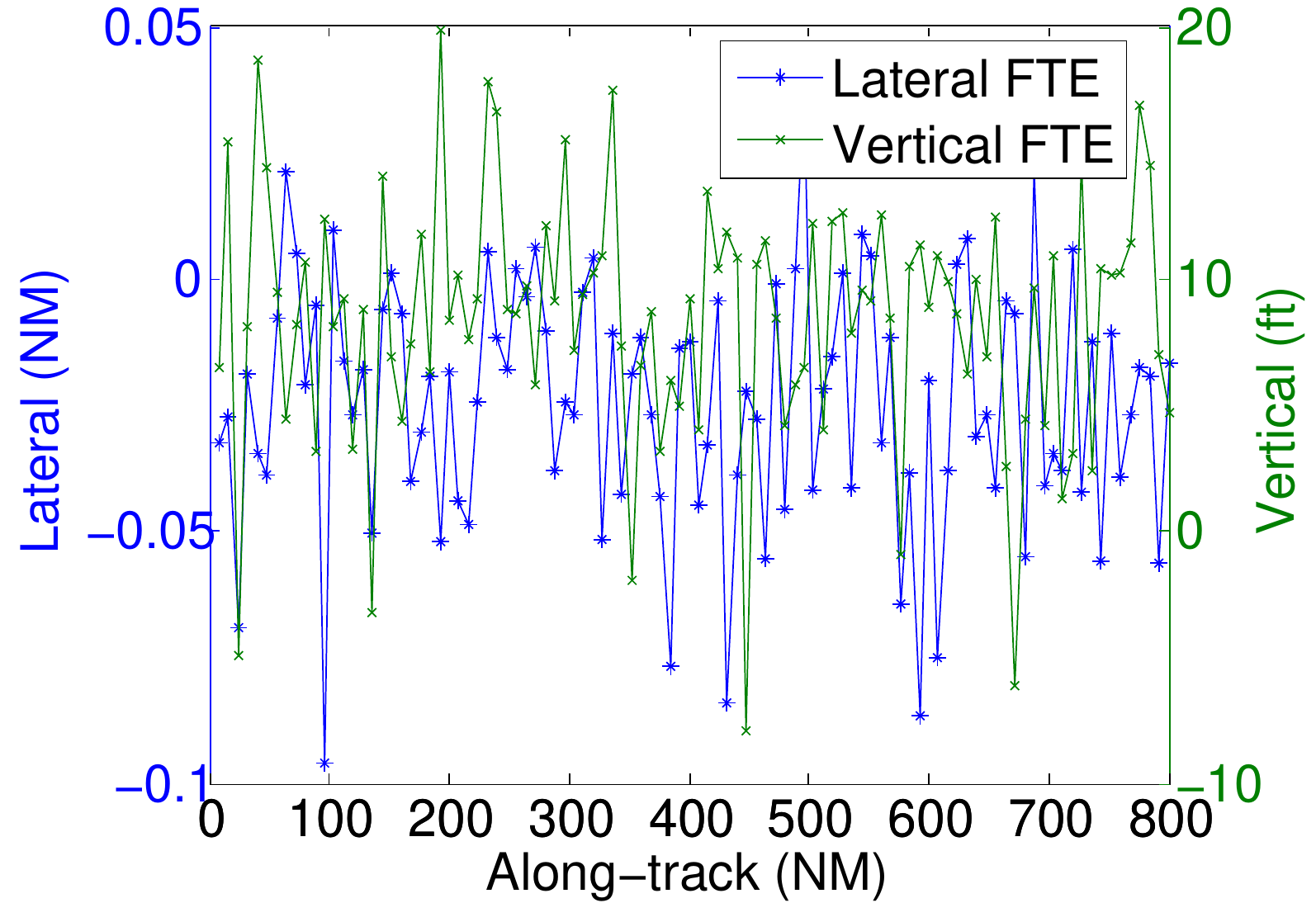}
\caption{Simulated FTE data}
\label{fig:FTEsim}
\end{figure}

\begin{figure}[!h]
\centering
\subfigure[Lateral]{
\includegraphics[width=1.6in]{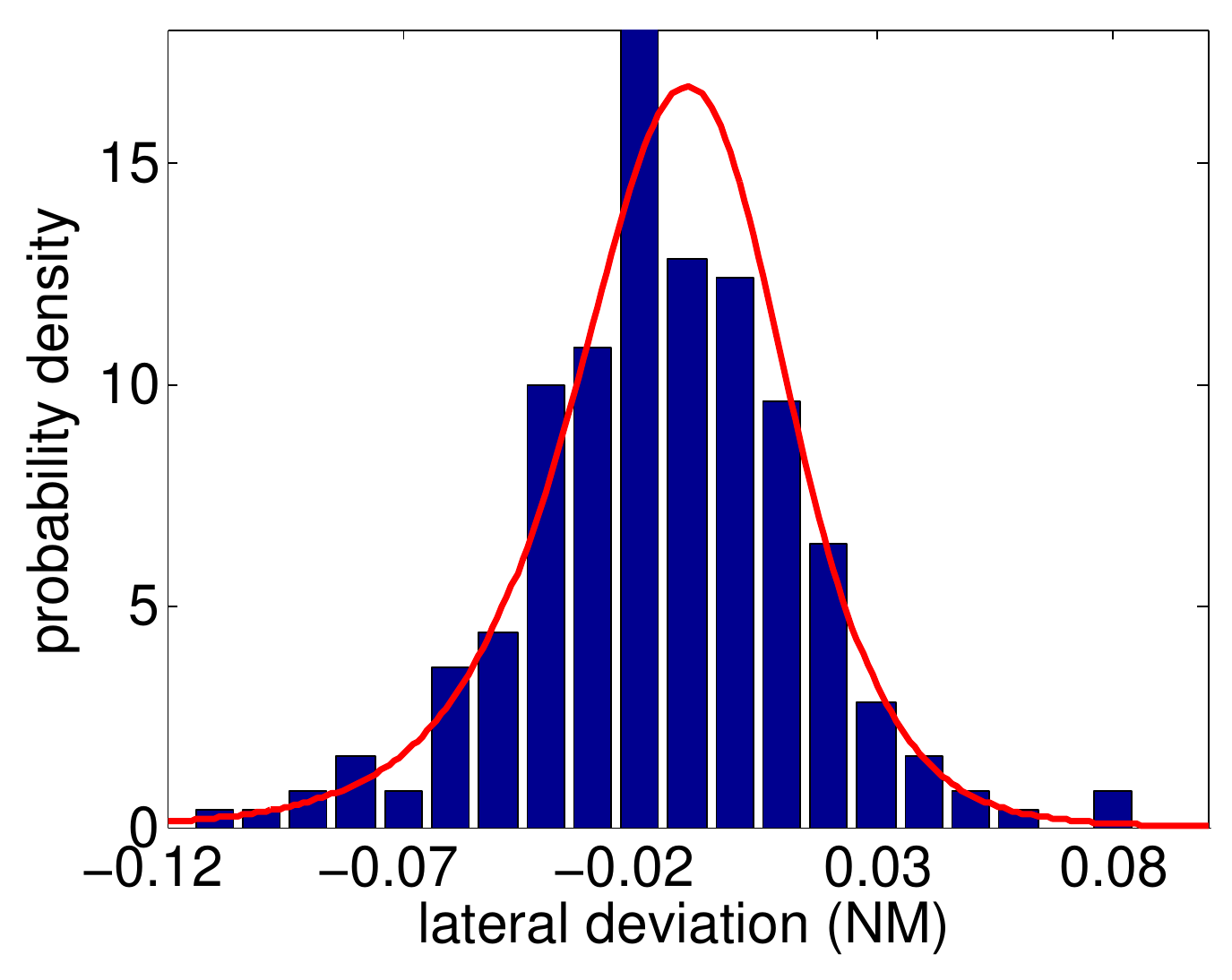}
}
\subfigure[Vertical]{
\includegraphics[width=1.6in]{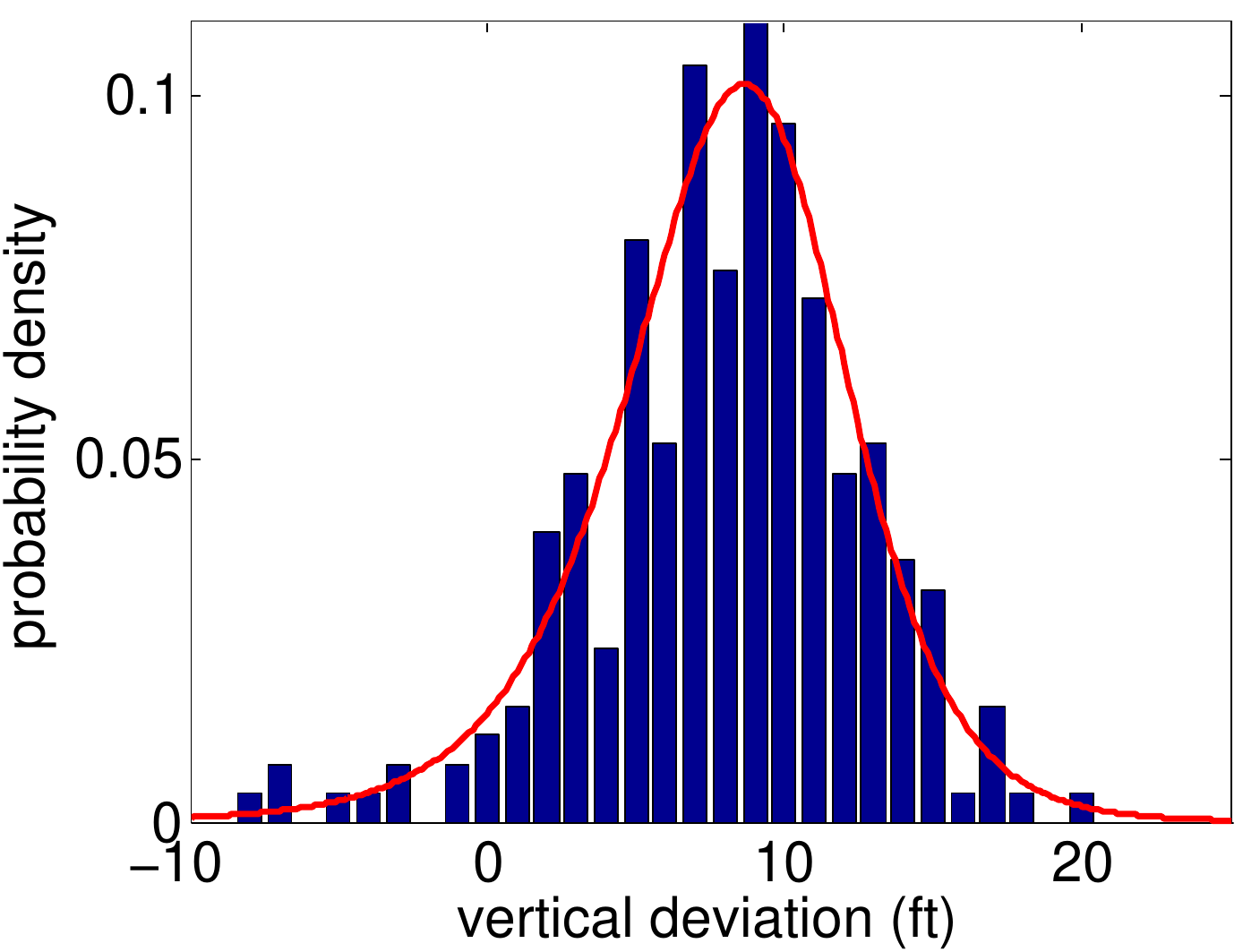}
}
\subfigure[Longitudinal]{
\includegraphics[width=1.6in]{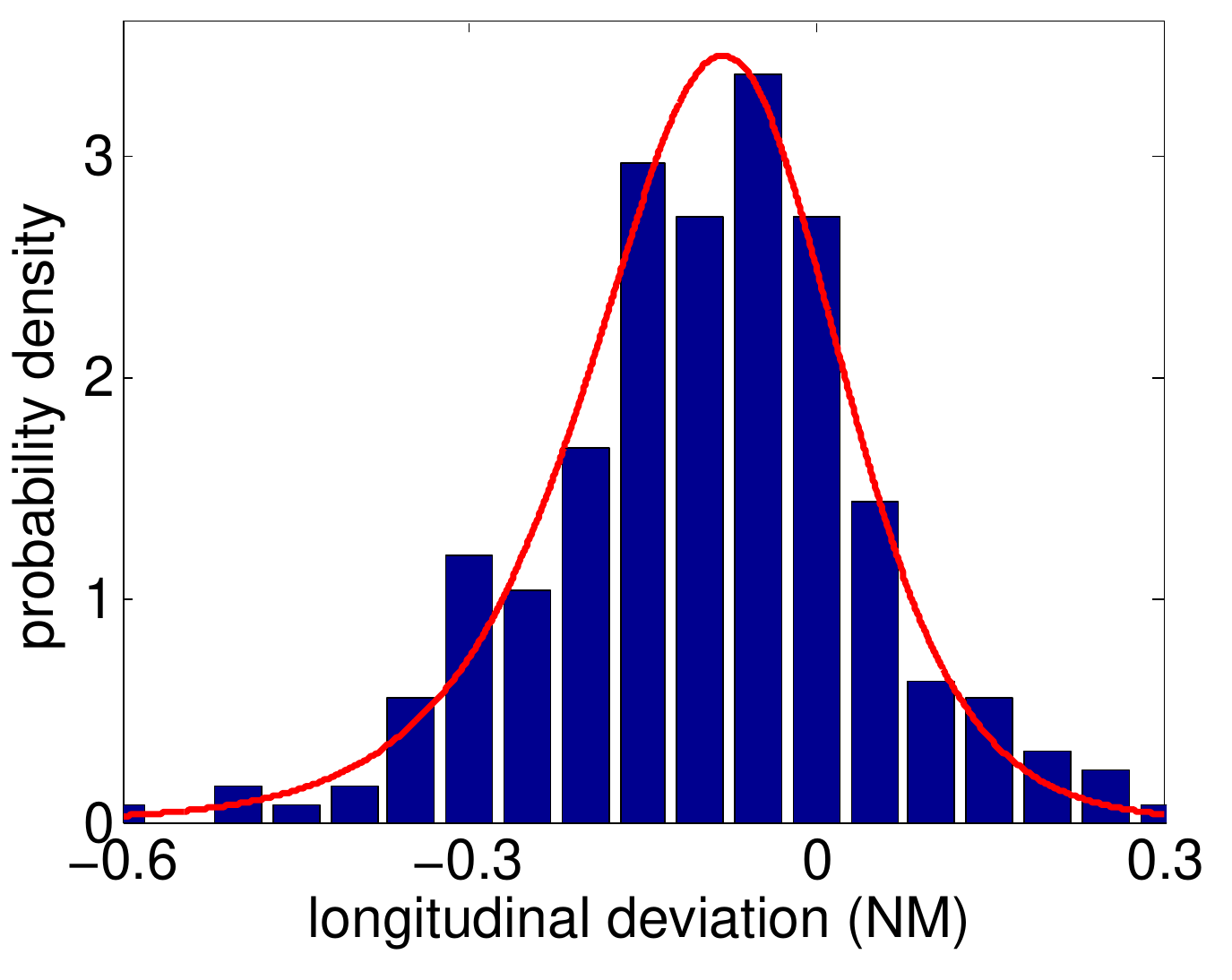}
}
\caption{Simulated FTE histograms and theoretical density functions}
\label{fig:FTEhist}
\end{figure}

Due to the practical difficulty in obtaining experimental FTE recordings, the numerical calibration of the Ornstein-Uhlenbeck model (\ref{eqn:OU}) for the aircraft was performed based on simulated data. In conformance to the findings of Levy et al., the Johnson unbound system $S_U$ in (\ref{eqn:JohnsonSU}) was used to generate fictitious FTE data (see Figures \ref{fig:FTEsim} and  \ref{fig:FTEhist}). The random number generator parametrized by the values in Table \ref{tab:FTEparam} provided data adhering to the moments in Table \ref{tab:FTEfit}, and the sampling time step for this data was assumed to be 1 minute.

By applying It\={o}'s lemma, it can be shown that for any fixed $s$ and $t$, $0\le s\le t$, the random variable $X_t$ conditional upon $X_s$ of the Ornstein-Uhlenbeck process has the form (\ref{eqn:OUproba}), where $\mathcal N(0,1)$ denotes a standard normal distribution. The relationship (\ref{eqn:OUrecursion}) between consecutive observations $X_i$ and $X_{i+1}$ with a timestep $\delta t$ is therefore affine with an independant and identically distributed random noise $\epsilon$.
\begin{align}
X_t&=X_s e^{-\kappa (t-s)}+\mu(1-e^{-\kappa (t-s)})\ldots\notag\\
&\ldots +\sigma\left[\frac{1-e^{-2\kappa (t-s)}}{2\kappa}\right]^{\frac{1}{2}}\cdot\mathcal N(0,1)\label{eqn:OUproba}\\
X_{i+1}&=aX_i+b+\epsilon \label{eqn:OUrecursion}
\end{align} 

A least squares linear regression is used to identify the recursion parameters $a$, $b$, and the standard deviation of the noise $\sigma_{\epsilon}$, from which the parameters (\ref{eqn:OUleastsq}) of the Ornstein-Uhlenbeck stochastic differential equation (\ref{eqn:OU}) can be deduced.
\begin{align}
\kappa &=-\frac{\ln a}{\delta t}\notag\\
\mu &=\frac{b}{1-a}\notag\\
\sigma &=\sigma_{\epsilon}\left[\frac{-2\ln a}{\delta t(1-a^2)}\right]^{\frac{1}{2}}\label{eqn:OUleastsq}
\end{align}

A maximum likelihood estimate method was also conducted. From (\ref{eqn:OUrecursion}), $\{X_{i+1}-aX_i-b=\epsilon\}$ is a normal random variable, and so the conditional probability density function of $X_{i+1}$ given $X_i$ with a time step $\delta t$ is shown in (\ref{eqn:OUcondprob}), while the log-likelihood function of $n+1$ observations $\{X_0,\cdots X_n\}$ is given in (\ref{eqn:OUloglike}).

\begin{align}
f_{[X_{i+1}|X_i]}(x)&=\frac{1}{\sqrt{2\pi\hat{\sigma}^2}}\ldots\notag\\
&\ldots\times\exp\left[-\frac{\left(x-X_i e^{-\kappa\delta t}-\mu(1-e^{-\kappa\delta t})\right)^2}{2\hat{\sigma}^2} \right]\label{eqn:OUcondprob}\\
\hat{\sigma}&=\sigma\left[\frac{1-e^{-2\kappa\delta t}}{2\kappa}\right]^{\frac{1}{2}}\notag\\
\mathcal L(\mu,\kappa,\hat{\sigma})&=\sum\limits_{i=0}^{n-1} \ln f_{[X_i+1|X_{i}]}(X_{i+1})\notag \\
\cdots&=-\frac{n}{2}\ln 2\pi - n\ln\hat{\sigma}\ldots\notag\\
&\ldots-\frac{1}{2\hat{\sigma}^2}\sum\limits_{i=0}^{n-1} [X_{i+1}-X_i e^{-\kappa\delta t}-\mu(1-e^{-\kappa\delta t})]^2 \label{eqn:OUloglike}
\end{align}

The argument of the maximum of $\mathcal L$ found from the three partial derivatives $\frac{\partial\mathcal L}{\partial \kappa}$, $\frac{\partial\mathcal L}{\partial \mu}$, $\frac{\partial\mathcal L}{\partial \hat{\sigma}}$ gives the system in (\ref{eqn:OUmaxlike}).

\begin{align}
\kappa&=-\frac{1}{\delta t}\ln\frac{\sum\limits_{i=0}^{n-1}(X_{i+1}-\mu)(X_i-\mu)}{\sum\limits_{i=0}^{n-1}(X_{i+1}-\mu)^2}\notag \\
\mu&=\frac{\sum\limits_{i=0}^{n-1} X_{i+1}-X_i e^{-\kappa\delta t}}{n(1-e^{-\kappa\delta t})}\notag\\
\hat{\sigma}^2&=\frac{1}{n}\sum\limits_{i=0}^{n-1}[X_{i+1}-\mu e^{-\kappa\delta t}(X_i-\mu)]^2 \label{eqn:OUmaxlike}
\end{align}

The resulting parameters $\kappa$, $\mu$, $\sigma$ of the Ornstein-Uhlenbeck model - for spatial units expressed in nautical miles (lateral and longitudinal) or feet (vertical) and temporal units expressed in minutes - are shown in Table \ref{tab:OUparam}. These values coincide for both the maximum likelihood and least squares estimation methods.

\begin{table}
\caption{Ornstein-Uhlenbeck aircraft model parameters}
\label{tab:OUparam}
\centering
\begin{tabular}{|c|c|c|c|}
\hline\hline
&Lateral&Vertical&Longitudinal \bigstrut\\ \hline\hline
$\kappa$ & 3.492 min$^{-1}$& 1.841 min$^{-1}$&2.1662 min$^{-1}$ \bigstrut \\ \hline
$\mu$ & $2.79\cdot 10^{-2}$ NM & 8.034 ft& $9.965\cdot 10^{-2}$ NM \bigstrut \\ \hline
$\sigma$ & $7.27\cdot 10^{-2}$ NM& 8.683 ft& 0.2774 NM \bigstrut\\ \hline \hline
\end{tabular}
\end{table}

\subsection{Flow}
We adopt a widely accepted model of interarrival times distributed according to an invariant Poisson process; this is equivalent to modeling passage times along the flow by an exponential distribution, and is a fair approximation of the system behavior over a short-time horizon (hourly time frame). If $N(t)$ is a cumulative count of all aircraft having entered the sector, then the probability distribution for the number of aircraft entering a flow between times $t$ and $t+\tau$ is given by (\ref{eqn:Poiss}) and the probability density for the values of interarrival times $x$ is given by (\ref{eqn:exp}).

\begin{align}
&\mathcal P[N(t+\tau)-N(t)=k]=\frac{e^{-\lambda\tau}(\lambda\tau)^k}{k!}\label{eqn:Poiss}\\
&f_{\lambda}(x)=\lambda\exp^{-\lambda x}, x\ge 0 \label{eqn:exp}
\end{align}

Although most rigorously, the distribution of velocities inside a flow is also a random variable \cite{Sal11}, we will assume the uniformity of velocities at 480 knots. This can be justified in the context of 4-D trajectories by constraints added to the along-track performance of the aircraft, which is not the case in current procedures. For a more realistic case, the flow would need to be modeled as a \emph{birth-death} process with two different Poisson processes, on entry and on exit from the controlled region. This adds some difficulty to the problem, without making it prohibitive.

The statistical study of the typical traffic entering Cleveland center conducted by Sala\"{u}n et al. \cite{Sal10} provides mean aircraft spacing between 50 and 200 NM for the ten busiest flows in the sector, and an hourly Poisson intensity parameter $\lambda_{center}$ between 10 and 100 (aircraft per hour) depending on time of day, with values averaging 80 for most of the busy times (6 AM to 8 PM EST). For the ten busiest flows in the center, the velocity and spacing values provide average intensity parameters $\lambda_{flow}$ from 10 to 2.5 (aircraft per hour). A reasonable estimate for the time an aircraft spends in a sector is 20 minutes.

\begin{figure}[!h]
\centering
\includegraphics[width=1.6in]{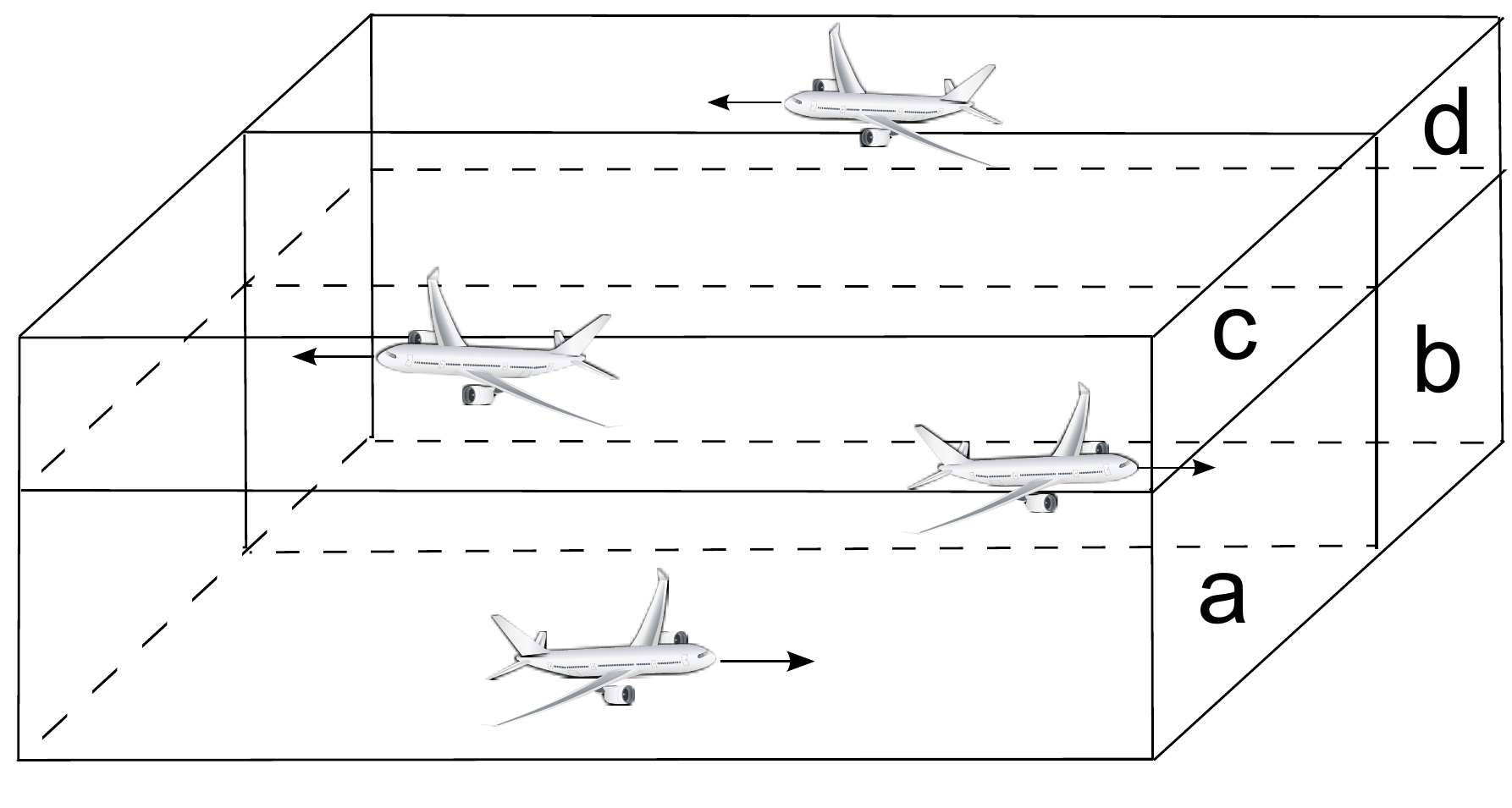}
\caption{Multilane flow corridors}
\label{fig:multiflow}
\end{figure}

From a spatial perspective, current en route flows have wide boundaries, reaching up to 40 NM in width, or even defying any reasonably defined geographic structure in some cases; we propose that the newly designed flow corridors will concentrate traffic to much tighter regions, with corridor bounds ranging between 3 NM and 10 NM lateral, and 1000-2000 ft vertical, and also supporting parallel lanes, or stacked corridors (see Figure \ref{fig:multiflow}). The flow model parameters are shown in Table \ref{tab:flow_param}.

\begin{table}
\caption{RNP corridor model parameters}
\label{tab:flow_param}
\centering
\begin{tabular}{|l|ll|}
\hline\hline
Intensity (a.c./h) 			& $\lambda_{inf}=2.5$  	&`$\lambda_{sup}=10$ \bigstrut\\ \hline
Lateral bounds (NM) 		& $y_{inf}=0.6$ 				& $y_{sup}=1$ \bigstrut \\ \hline
Vertical bounds (ft) 		& $z_{inf}=60$ 					& $z_{sup}=120$\bigstrut \\ \hline
Longitudinal bounds (NM)& $x_{inf}=1$ 					& $x_{sup}=5$ \bigstrut\\ \hline\hline
\end{tabular}
\end{table}

\section{Expression of taskload probabilities}
\label{sec:analytic}

In this section we consider several basic functional elements of the flow corridor structure.

\subsection{Single lane}
\label{sec:1lane}
In the case of a single lane of traffic, the overall controller taskload is a superposition of the individual effects from each aircraft. From the Poisson process flow model in (\ref{eqn:Poiss}), the probability of the number of aircraft $M$ simultaneously controlled (a random variable) can be deduced: if it takes an aircraft $T_{cross}$ to cross the airspace, then the probability that at any given time there are $k$ aircraft present is given in (\ref{eqn:nb_ac}); if there are $M=k\ge1$ aircraft present, and the aircraft $i$ requires $N_i^{(T)}$ interventions over time $T$, then the total taskload the controller is subject to has a probability given in (\ref{eqn:k_ac_task}).
\begin{align}
\mathcal P[M=k]&=\frac{e^{-\lambda T_{cross}}(\lambda T_{cross})^k}{k!}\label{eqn:nb_ac}\\
\mathcal P[N_1^{(T)}+\cdots+N_k^{(T)}=n]&= \left[\stackrel{k-1}{\star}\right]\mathcal P\{N^{(T)}\}[n]\label{eqn:k_ac_task}
\end{align}

Here $\left[\stackrel{k-1}{\star}\right]\mathcal P\{N^{(T)}\}$ represents $k-1$ discrete autoconvolutions of the single aircraft taskload probability $\mathcal P\{N^{(T)}\}$ from (\ref{eqn:ProbFtask}). Obviously for $M=0$, there will be no intervention with probability 1.

\begin{align*}
\left[\stackrel{0}{\star}\right]\mathcal P\{N^{(T)}\}[n]&=\mathcal P[N^{(T)}=n] \\
\left[\stackrel{1}{\star}\right]\mathcal P\{N^{(T)}\}[n]&=\sum\limits_{i=0}^n \mathcal P[N^{(T)}=i]\cdot \mathcal P[N^{(T)}=n-i]\\
\left[\stackrel{m+1}{\star}\right]\mathcal P\{N^{(T)}\}[n]&=\sum\limits_{i=0}^n \left[\stackrel{m}{\star}\right]\mathcal P\{N^{(T)}\}[i]\cdot \mathcal P[N^{(T)}=n-i]\\
\end{align*}

Since the number of aircraft is a random variable, combining (\ref{eqn:nb_ac}) and (\ref{eqn:k_ac_task}) gives the probability (\ref{eqn:flow_task}) of the overall taskload $N^{(T)}_{\lambda}$ for the flow with intensity $\lambda$ over time $T$ for $n\ge1$; $\mathcal P[N^{(T)}=0]$ is the probability that no intervention is required for one aircraft, given by (\ref{ProbOtask_ac}).

\begin{align}
\mathcal P[N^{(T)}_{\lambda}=n]&=\sum\limits_{i=1}^{+\infty}\mathcal P[M=i]\cdot\left[\stackrel{i-1}{\star}\right]\mathcal P\{N^{(T)}\}[n] \label{eqn:flow_task}\\
\mathcal P[N^{(T)}_{\lambda}=0]&=\sum\limits_{i=0}^{+\infty}\mathcal P[M=i]\cdot \left(\mathcal P[N^{(T)}=0]\right)^i\notag\
\end{align}

\subsection{Multiple parallel lanes}
\label{sec:multilane_model}
Let us consider the case of parallel lanes (see Figure \ref{fig:multiflow}), with either opposing or matching traffic orientation.  If the spatial extents are identical, and assuming independence of the flows, the whole system is equivalent to a single Poisson process with cumulative intensities $\lambda_{tot}=\sum\lambda_i$. Results from Section \ref{sec:1lane} can thus be simply generalized to (\ref{eqn:multi_flow_task}).

\begin{align}
\mathcal P\{N^{(T)}_{\cup\lambda_i}\}= \mathcal P\{N^{(T)}_{\sum\lambda_i}\} \label{eqn:multi_flow_task}
\end{align}

If the spatial extents of the flows are different, the problem is slightly more complex. The taskload probability (\ref{eqn:multi_flow_diff_ext_task}) for $j\ge2$ different flows with $\{e_1, \cdots, e_j\}$ spatial extents, $\{\lambda_1, \cdots, \lambda_j\}$ intensities, and $\{N_{\lambda_1,e_1}^{(T)},\cdots, N_{\lambda_j,e_j}^{(T)}\}$ interventions per flow is

\begin{align}
\mathcal P[N^{(T)}_{\cup\lambda_i}=n]&= \mathcal P[N_{\lambda_1,e_1}^{(T)}+\cdots+N_{\lambda_j,e_j}^{(T)}=n]\notag\\
\cdots&=\left[\displaystyle\mathop{\star}\limits_{i=1}^j\right]\mathcal P\{N_{\lambda_i,e_i}^{(T)}\}[n]	\label{eqn:multi_flow_diff_ext_task}
\end{align}

where $\left[\displaystyle\mathop{\star}\limits_{i=1}^j\right]$ designates the successive convolutions of the $\mathcal P\{N_{\lambda_i,e_i}^{(T)}\}$ taskload probabilities given in (\ref{eqn:flow_task}) for flows 1 to $j$. This operator is well defined since convolution is associative.

\subsection{Crossings and mergings}

\begin{figure}[!h]
\centering
\includegraphics[width=1.6in]{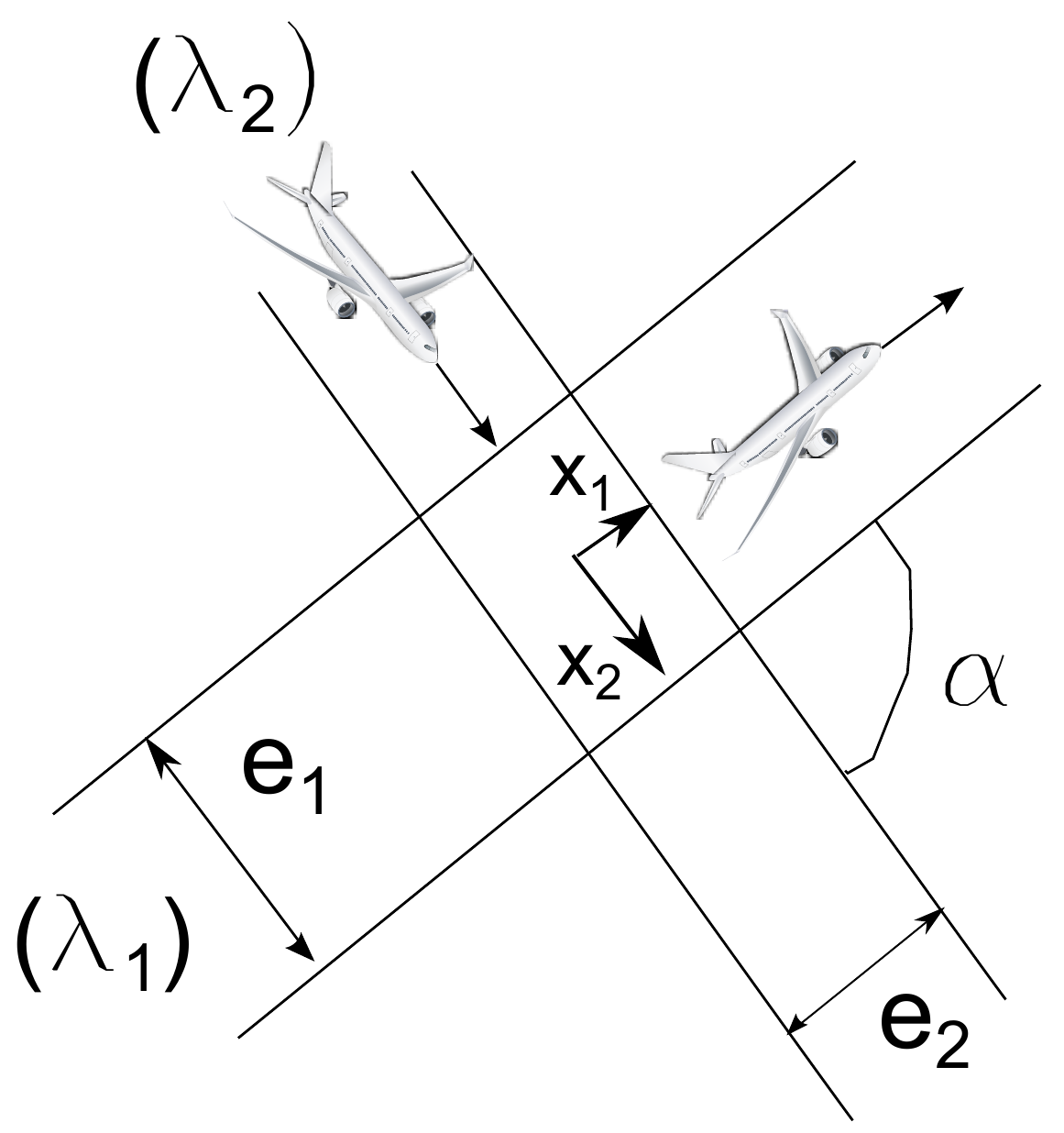}
\caption{Crossing of flows}
\label{fig:cross}
\end{figure}

\begin{figure}[!h]
\centering
\subfigure[$x_1$ bounds]{
\includegraphics[width=1.6in]{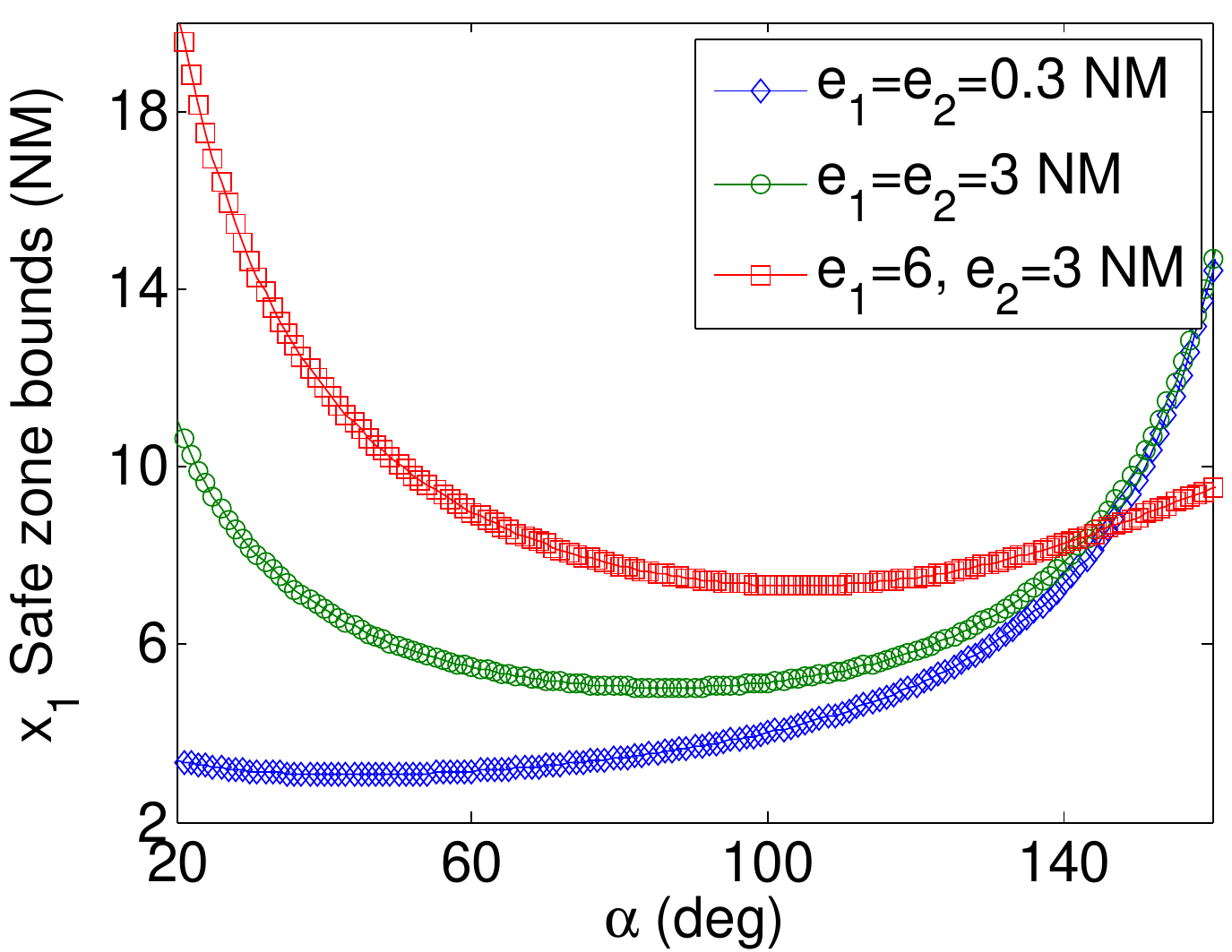}
}
\subfigure[$x_2$ bounds]{
\includegraphics[width=1.6in]{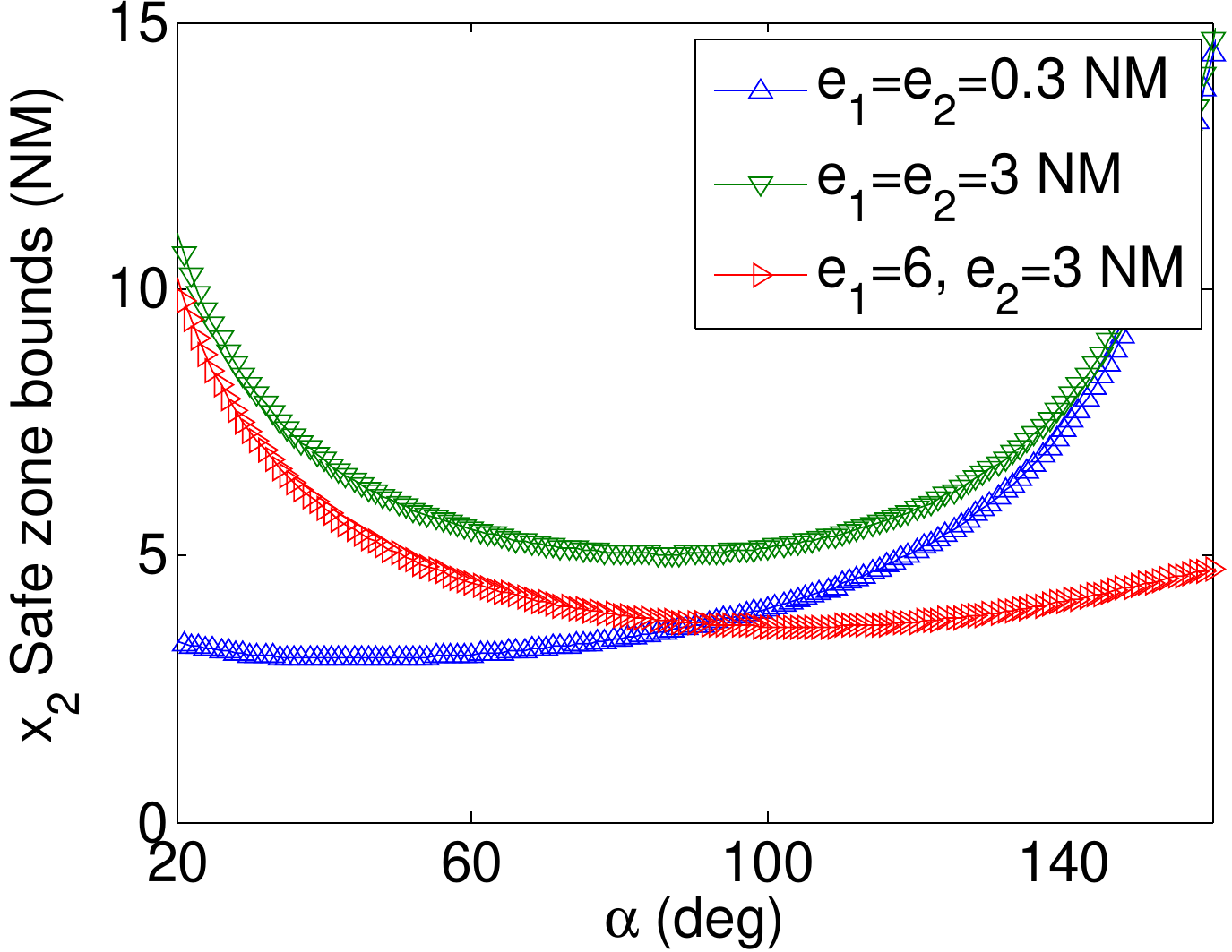}
}
\caption{Possible safe-zone bounds solutions}
\label{fig:safe_bounds}
\end{figure}

In the case of a crossing, an additional source of taskload comes from possible scheduling conflicts. For two flows $\{F_1\}$ and $\{F_2\}$ crossing at an angle $\alpha$ (in the plane defined by the centerlines of these two flows, see Figure \ref{fig:cross}) with lateral extents $e_1$ and $e_2$, Poisson intensities $\lambda_1$ and $\lambda_2$, a \emph{safe-zone} can be defined around the intersection of the centerlines, where only one aircraft is allowed simultaneously. For a minimum possible approach distance $D_{min}$ (taken to be for example 5 NM, i.e. the conflict separation standard), the symmetrical safe-zone boundaries $x_1$ and $x_2$ in each of the flows are defined by (\ref{eqn:safe_bounds}). Figure \ref{fig:safe_bounds} shows plots of some of these boundary values.

\begin{align}
(x_1&+x_2\cos\alpha -\frac{e_2}{2}\cos\alpha )\ldots \notag \\
&\ldots+(-\frac{e_1}{2}+x_2\sin\alpha+\frac{e_2}{2}\sin\alpha )^2 = D_{min}^2\notag\\
(x_1&+x_2\cos\alpha -\frac{e_2}{2}\sin\alpha)^2\ldots\notag \\
&\ldots+(-\frac{e_1}{2}+x_2\sin\alpha+\frac{e_2}{2}\cos\alpha)^2 = D_{min}^2 \label{eqn:safe_bounds}
\end{align}

The safe-zone can then be extended for simplicity so that its crossing time is equal along both flows. The probability of scheduling conflicts is once again simply found from superposing the Poisson processes: if passing the safe-zone requires a time $T_{safe}$, then the probability that there are $A=c$ aircraft simultaneously present inside the safe-zone is (\ref{eqn:nb_conf}).

\begin{align}
\mathcal P[A=c]=\frac{e^{-(\lambda_1+\lambda_2) T_{safe}}[(\lambda_1+\lambda_2) T_{safe}]^c}{c!}\label{eqn:nb_conf}
\end{align}

This can also be interpreted as a \emph{lower bound} of the number of required conflict resolution interventions: $c-1$ of those aircraft will need to be delayed and/or deviated, without accounting for possible secondary conflicts. This probability however does account for scheduling conflicts within each separate flow, as well as the crossing flows combined.

From (\ref{eqn:flow_task}), (\ref{eqn:multi_flow_task}), and (\ref{eqn:nb_conf}), a discrete convolution $[\star]$ in (\ref{eqn:cros_task}) gives the total taskload over time $T$ for $n\ge1$ accounting for $A-1$ conflicts at the flow crossing and for $N^{(T)}_{\lambda_1\cup\lambda_2}$ interventions to maintain the structure.

\begin{align}
\mathcal P[N_{\lambda_1\times\lambda_2}^{(T)}=n]&=\mathcal P[A-1+N_{\lambda_1\cup\lambda_2}^{(T)}=n]\notag \\
\cdots &=\mathcal P\{A-1\}[\star]\mathcal P\{N^{(T)}_{\lambda_1\cup\lambda_2}\}\notag \\
\cdots &=\sum\limits_{i=0}^n \mathcal P[A=i+1]\cdot \mathcal P[N^{(T)}_{\lambda_1\cup\lambda_2}=n-i]\label{eqn:cros_task}\\
\mathcal P[N_{\lambda_1\times\lambda_2}^{(T)}=0]&=\mathcal P[A=0]+\mathcal P[A=1]\cdot\mathcal P[N^{(T)}_{\lambda_1\cup\lambda_2}=0] \notag
\end{align}

\section{Monte Carlo simulation}
\label{sec:sim}

Several Monte Carlo simulations were conducted in order to estimate the resulting taskload. Four control tolerance standards (\emph{stringent}, \emph{severe}, \emph{intermediate}, \emph{lax}) were defined, shown in Table \ref{tab:1flow_tol}. The control bounds are extremely restrictive in order to illustrate the low probabilities of oscillation. The occurance of a lateral FTE of 0.3 NM has a probability of $5\cdot10^{-6}$ in the Ornstein-Uhlenbeck model used in the simulation, which corresponds to a probability of $6\cdot10^{-4}$ for one intervention required over 2 hours. For a lateral FTE of 0.4 NM, the probability drops below $2\cdot10^{-7}$, which is the resolution limit of the Monte Carlo simulation. For each run, the taskload calculation horizon was $T=2$ hours.

\begin{table}
\caption{Tolerance standards}
\label{tab:1flow_tol}
\centering
\begin{tabular}{|c|c|c|c|}
\hline\hline 
 & Lateral (NM) & Vertical (ft) & Longitudinal (NM) \bigstrut \\ \hline\hline
\emph{Stringent} & 0.1 &  20 & 0.5 \bigstrut \\ \hline
\emph{Severe}& 0.12 & 22 & 0.6  \bigstrut \\ \hline
\emph{Intermediate}& 0.15 & 25 & 0.8 \bigstrut \\ \hline
\emph{Lax} & 0.2 & 30 & 1 \bigstrut\\ \hline \hline
\end{tabular}
\end{table}

\subsection{Single lane}

\begin{figure}[!h]
\centering
\subfigure[Taskload probability for \emph{stringent} control bounds]{
\includegraphics[width=1.6in]{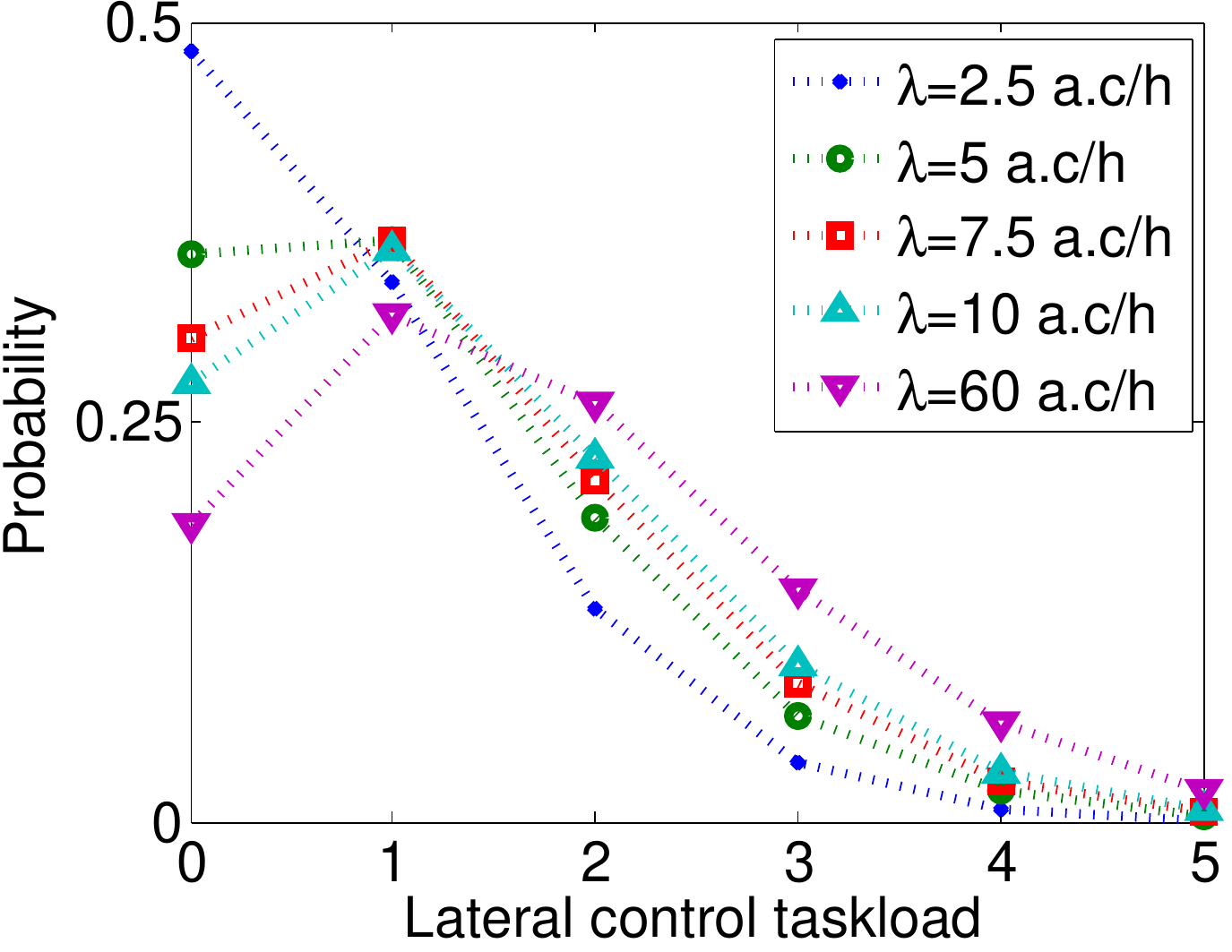}
}
\subfigure[Taskload for very high density flow ($\lambda~=~60$~a.c/h)]{
\includegraphics[width=1.6in]{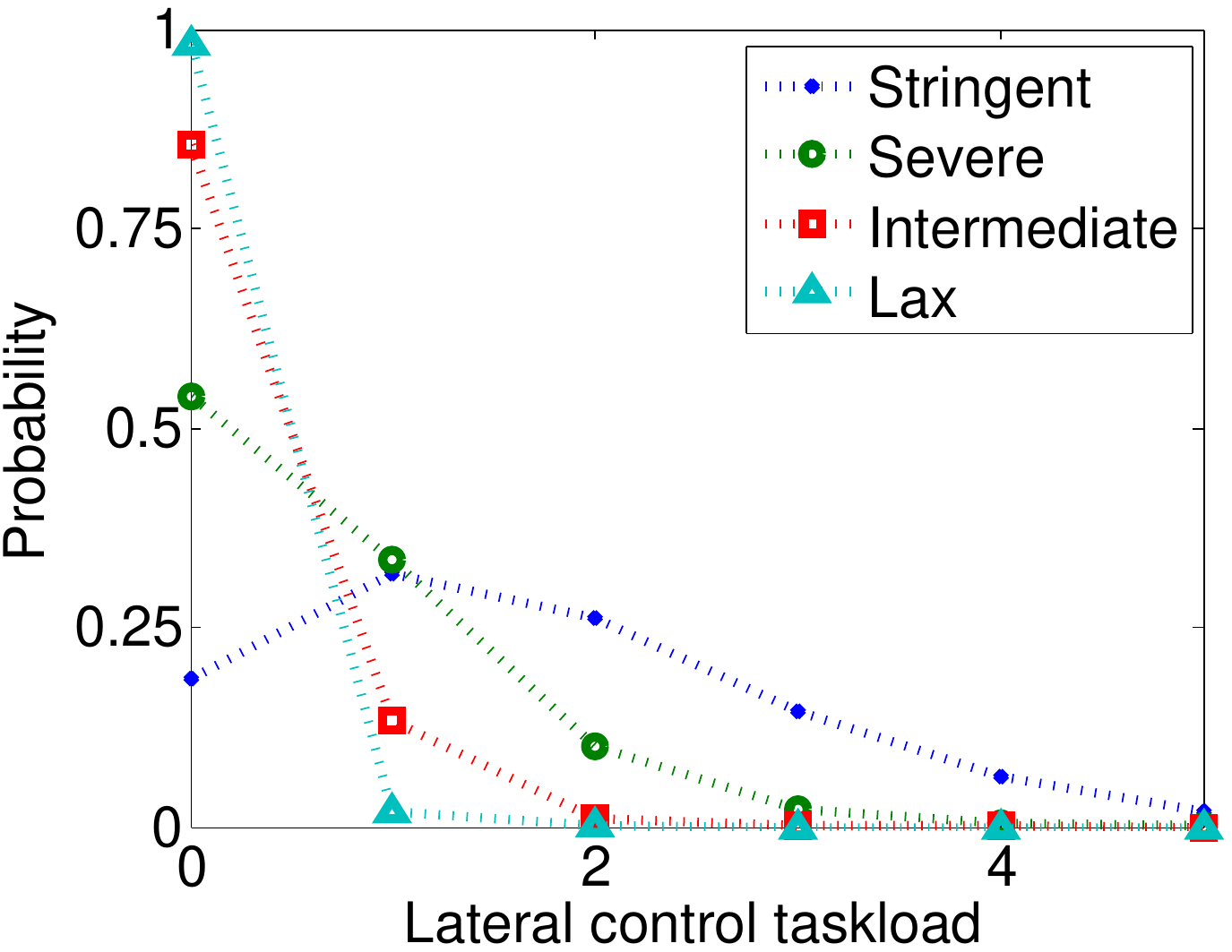}
}
\caption{Effect of flow rate and control tolerance on monolane lateral control 2h taskload probability}
\label{fig:task_lat_1fl}
\end{figure}

\begin{figure}[!h]
\centering
\subfigure[Taskload probability for \emph{stringent} control bounds]{
\includegraphics[width=1.6in]{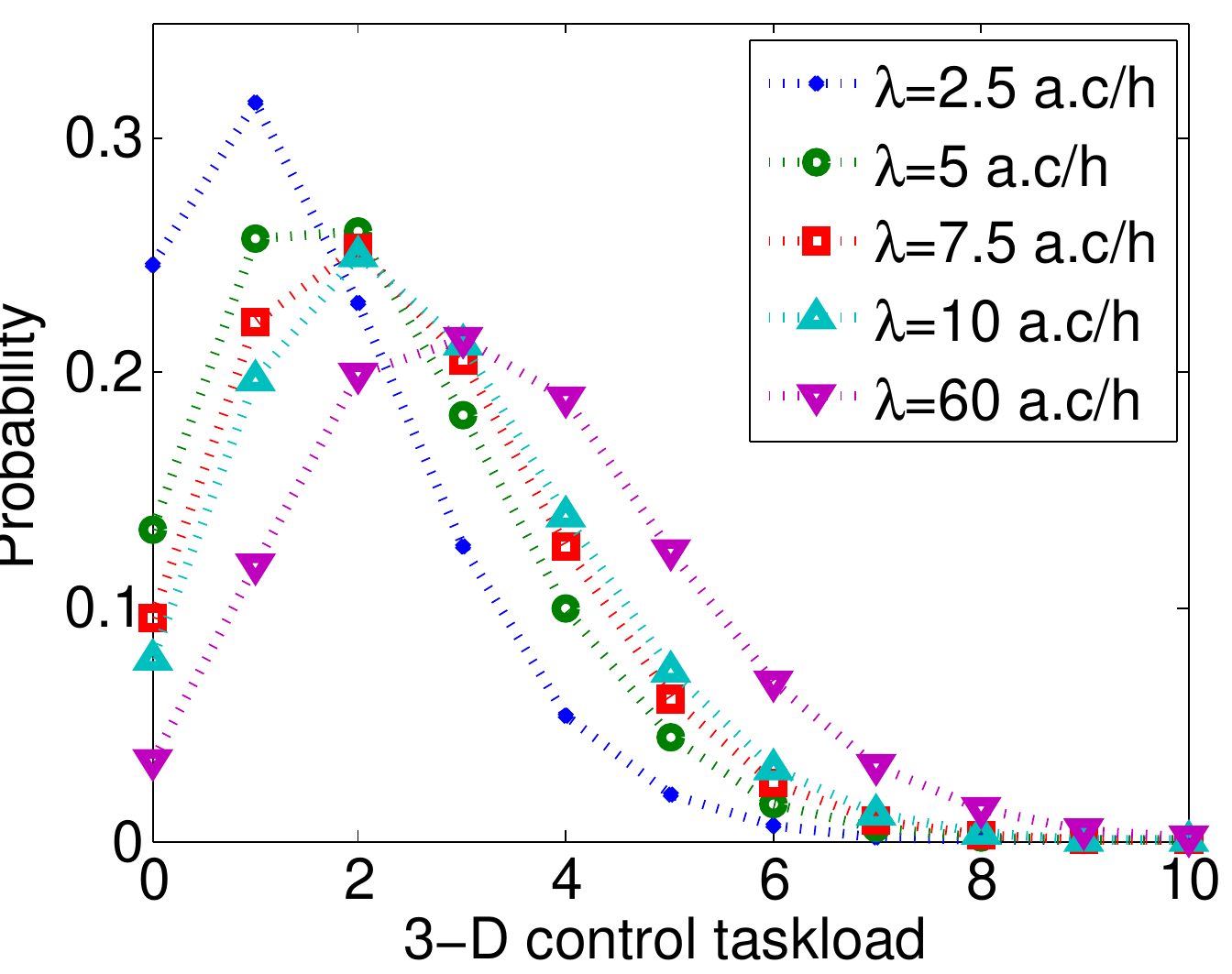}
}
\subfigure[Taskload probability for very high density flow ($\lambda~=~60$~a.c/h)]{
\includegraphics[width=1.6in]{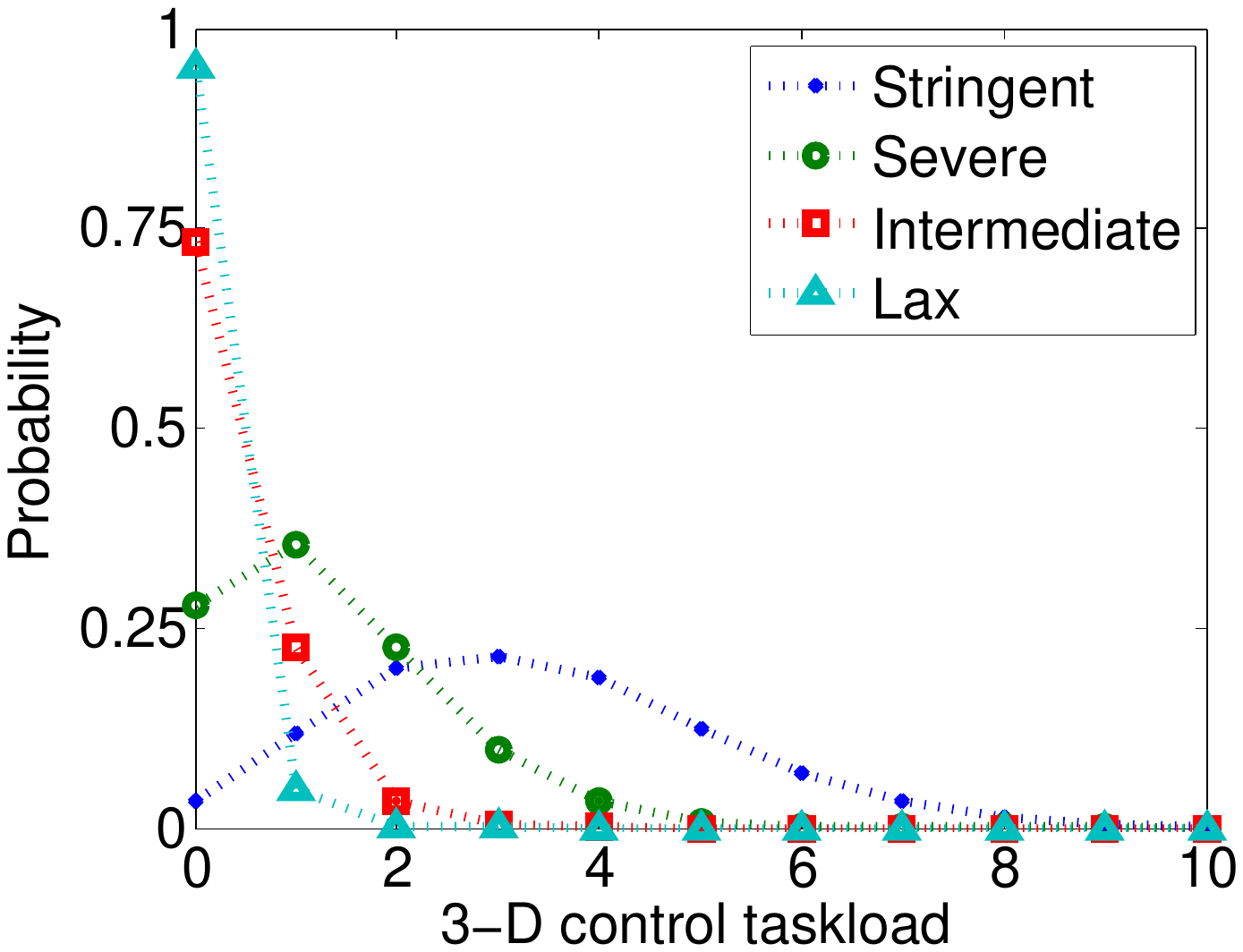}
}
\caption{Effect of flow rate and control tolerance on monolane 3-D (lateral, vertical, longitudinal) control 2h taskload probability}
\label{fig:task_tot_1fl}
\end{figure}

Several flow intensities were chosen and a varying number of runs were performed (detailed in Table \ref{tab:1flow_MC}) in order to simulate $2.5\cdot10^5$ aircraft for each flow intensity. The lateral containment taskload probability for a two hour interval is shown in Figure \ref{fig:task_lat_1fl}, and the cumulated lateral, vertical, and longitudinal taskload (3-D control) appears in Figure \ref{fig:task_tot_1fl}. As expected, the plots show increasing taskload with higher density flows and with stricter tolerance standards. Nevertheless, the values remain relatively low, with no more than 9 or 10 interventions being expected over a 2 hour time frame in all cases. 

\begin{table}
\caption{Monolane MC simulation parameters}
\label{tab:1flow_MC}
\centering
\begin{tabular}{|c|c|c|c|c|c|}
\hline\hline 
Intensity $\lambda$ (a.c/h)&  2.5 & 5 & 7.5 & 10 & 60 \bigstrut\\ \hline
MC runs& 91,658 & 66,680 & 58,366 & 54,147 & 41,702 \bigstrut \\ \hline
Simulated aircraft & \multicolumn{5}{c|}{250,000} \bigstrut \\ \hline\hline
\end{tabular}
\end{table}

\subsection{Multiple parallel lanes}

As noted in Section \ref{sec:multilane_model}, the case of several flows with identical spatial extents (or control tolerances) and varying Poisson intensities can be simply reduced to a single flow with cumulative intensity. The case of lanes with varying precision requirements poses more interest. In this simulation, the very high density flow intensity parameter (60 a.c/h) was chosen for all the lanes, while the control tolerance standards were \emph{stringent} for the first flow, \emph{severe} for the second, \emph{intermediate} for the third, and \emph{lax} for the fourth.

\begin{figure}[!h]
\centering
\subfigure[Lateral control taskload probability]{
\includegraphics[width=1.6in]{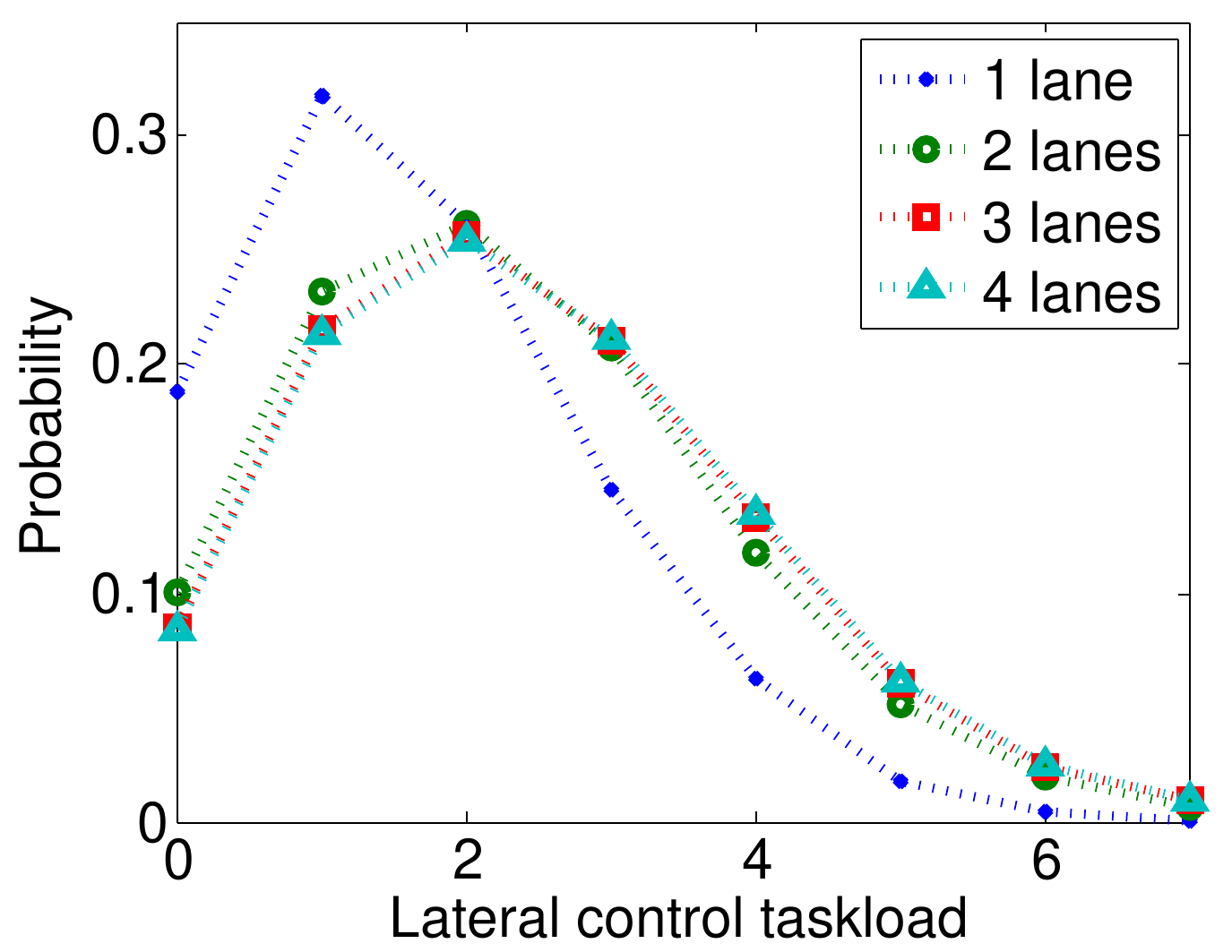}
}
\subfigure[3-D control taskload probability]{
\includegraphics[width=1.6in]{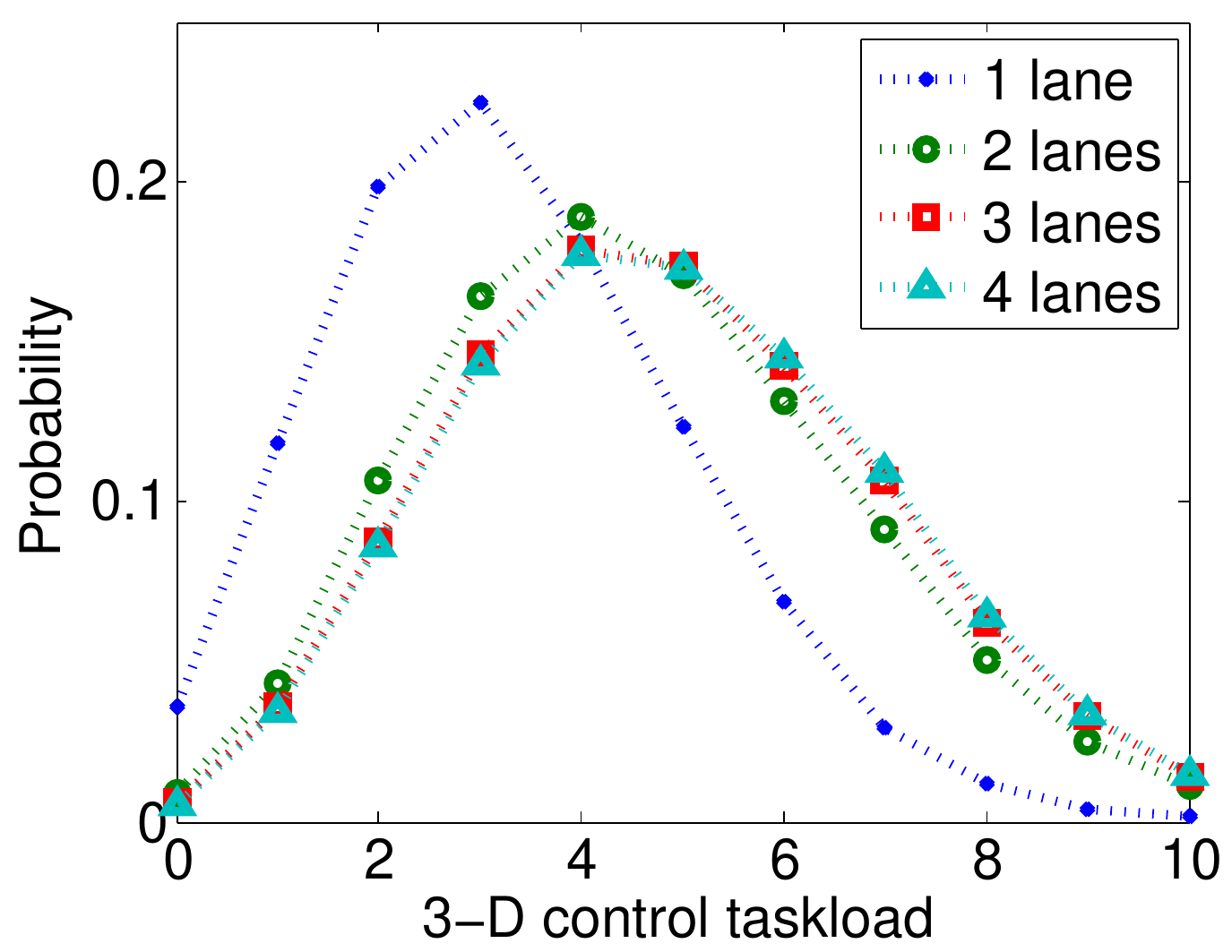}
}
\caption{Effect of number of lanes on multilane 2h taskload probability}
\label{fig:task_lat_multifl}
\end{figure}

The Monte Carlo simulation parameters of the multilane flow are shown in Table \ref{tab:multiflow_MC}. The taskload probability for a two hour interval is shown in Figure \ref{fig:task_lat_multifl}. The simulation reveals that although taskload increases with additional lanes, the \emph{stringent} flow is the source of most of the taskload. Adding the \emph{severe} flow to this has significant impact by increasing the most probable taskload level from 3 to 4 interventions over a two hour time frame in the 3-D control case. Additional flows with more tolerant bounds do not have notable effect, only increasing taskload probabilities by less than 1\%.

\begin{table}
\caption{Multilane MC simulation parameters}
\label{tab:multiflow_MC}
\centering
\begin{tabular}{|c|c|c|c|c|c|}
\hline\hline
Intensity $\lambda$ (a.c/h)&  2.5 & 5 & 7.5 & 10 & 60 \bigstrut\\ \hline
MC runs& 22,9158 & 16,670 & 14,592 & 13,537 & 10,426 \bigstrut \\ \hline
Simulated aircraft & \multicolumn{5}{c|}{62,500} \bigstrut\\ \hline\hline
\end{tabular}
\end{table}

\subsection{Crossings and mergings}
The Monte Carlo simulation parameters of the crossing are shown in Table \ref{tab:cross_MC}. The taskload calculations were performed for two crossing flow intensities of $\lambda=2.5$ a.c/h. It was assumed that in the enroute domain, the higher density flow corridors would be designed for the express purpose of avoiding such crossings. Simulations show that in the higher density levels, the taskload comes almost exclusively from resolving scheduling conflicts at the crossing, rather than from controlling the aircraft deviations.

\begin{table}
\caption{Crossing MC simulation parameters}
\label{tab:cross_MC}
\centering
\begin{tabular}{|c|c|c|c|}
\hline\hline
Crossing angle $\alpha$ (deg)&  30 & 90 & 120 \bigstrut\\ \hline
Crossing time (min) & 2 & 1 & 3 \bigstrut \\ \hline
MC runs& 5,412 &10,463 & 3,826\bigstrut \\ \hline
Simulated aircraft & 25,000 & 50,000 & 16,666 \bigstrut\\ \hline\hline
\end{tabular}
\end{table}

\begin{figure}[!h]
\centering
\subfigure[Deviation control taskload probability]{
\includegraphics[width=1.6in]{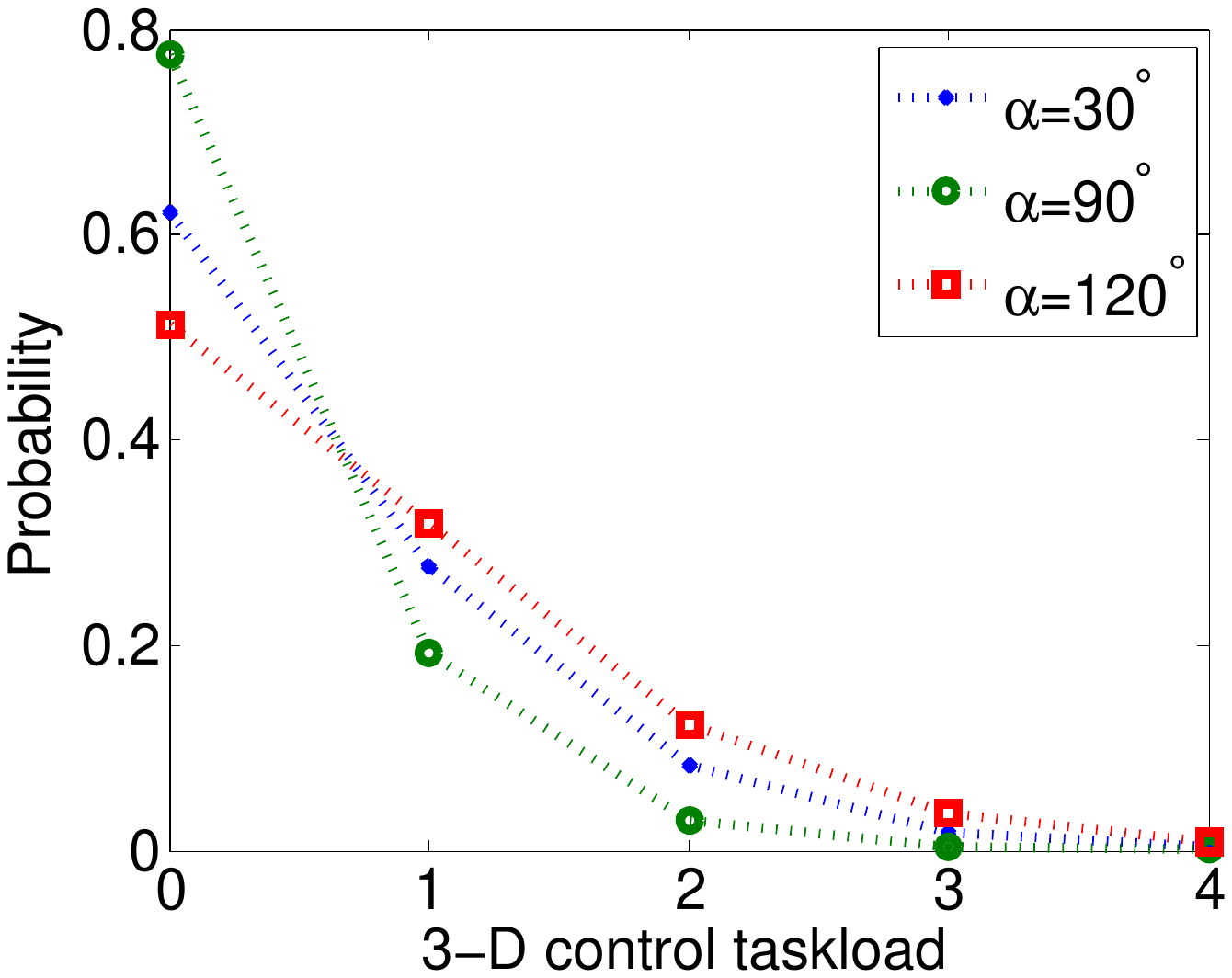}
}
\subfigure[Conflict resolution taskload probability]{
\includegraphics[width=1.6in]{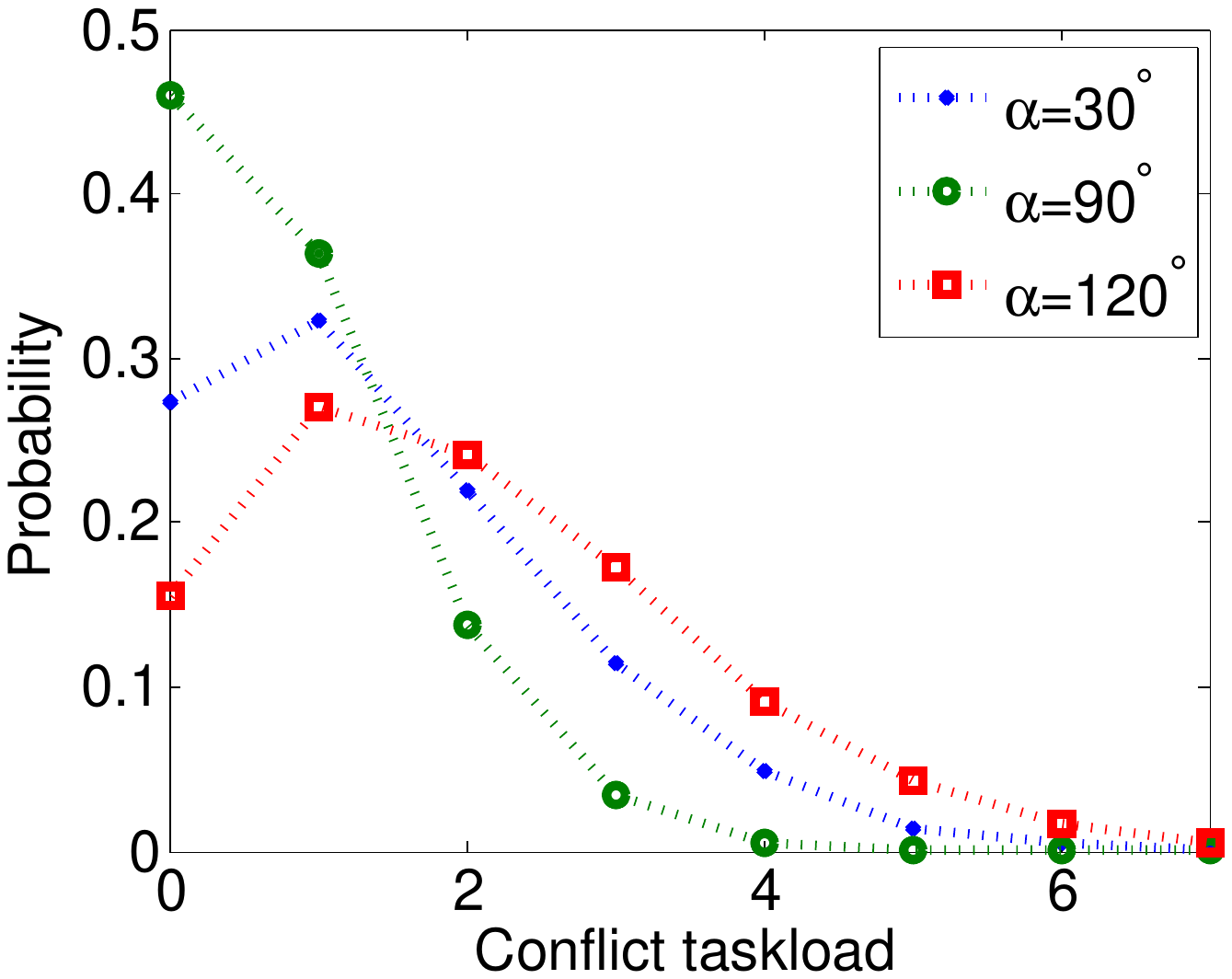}
}
\subfigure[Total taskload probability]{
\includegraphics[width=1.6in]{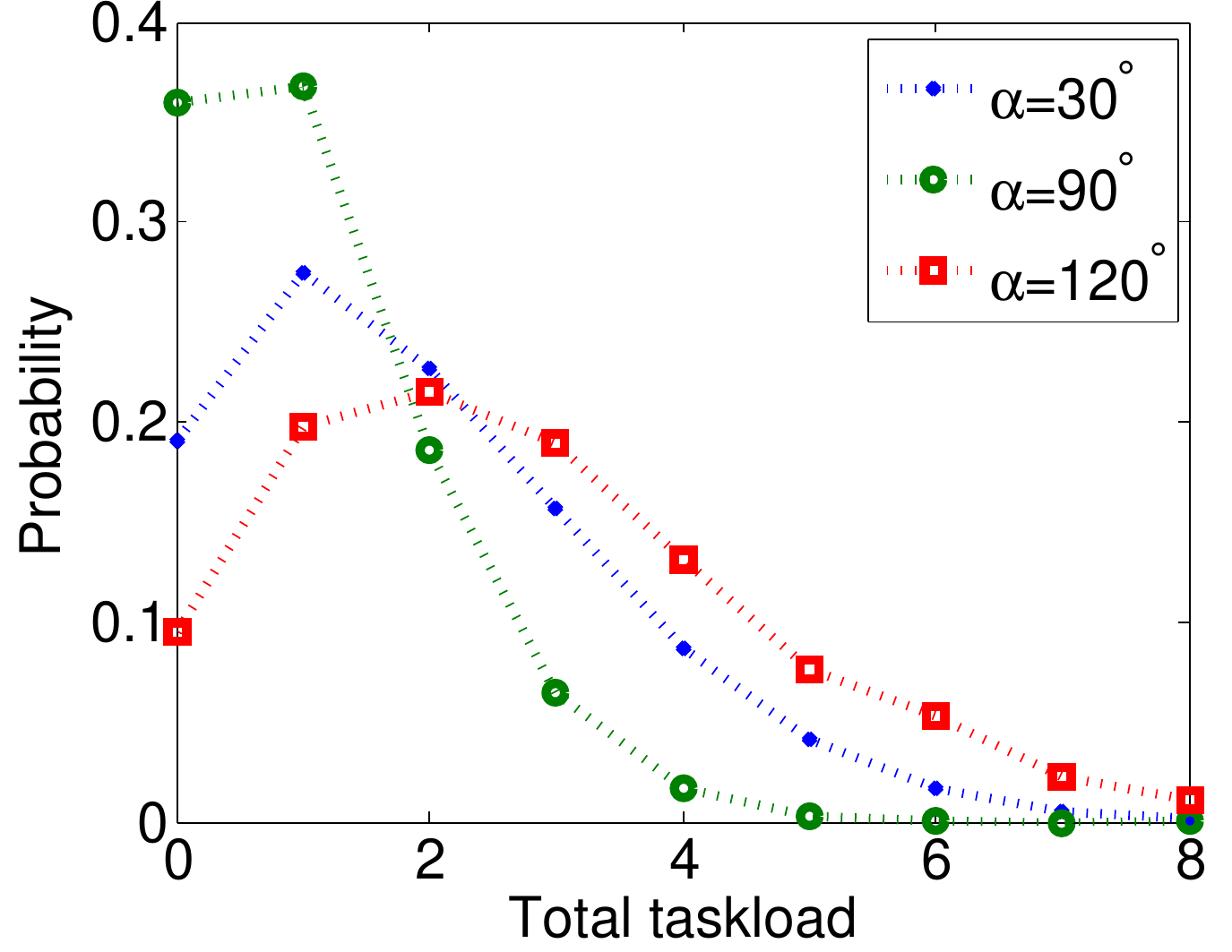}
}
\caption{Effect of crossing angle on 2h taskload probability; \emph{stringent} standards for both flows }
\label{fig:stringent_cross}
\end{figure}

Figure \ref{fig:stringent_cross} shows the effect of the crossing angle $\alpha$ on the controller taskload for a crossing of two flows with \emph{stringent} deviation tolerance. Since the $\alpha=90^{\circ}$ crossing angle imposes the smallest safe-zone extent, it it intuitive that this also corresponds to the lowest conflict resolution taskload, and inversely for the $\alpha=120^{\circ}$ crossing which causes the largest safe-zone. The spatial extent also explains the control taskload: the longer the time an aircraft spends within the the safe-zone, the higher the probability that it will touch one of the control bounds. Therefore, the $\alpha=90^{\circ}$ crossing has the highest probability that no control intervention will be needed, and the $\alpha=120^{\circ}$ crossing has the highest probabilities for any of the superior taskload values.

\begin{figure}[!h]
\centering
\subfigure[Deviation control taskload probability]{
\includegraphics[width=1.6in]{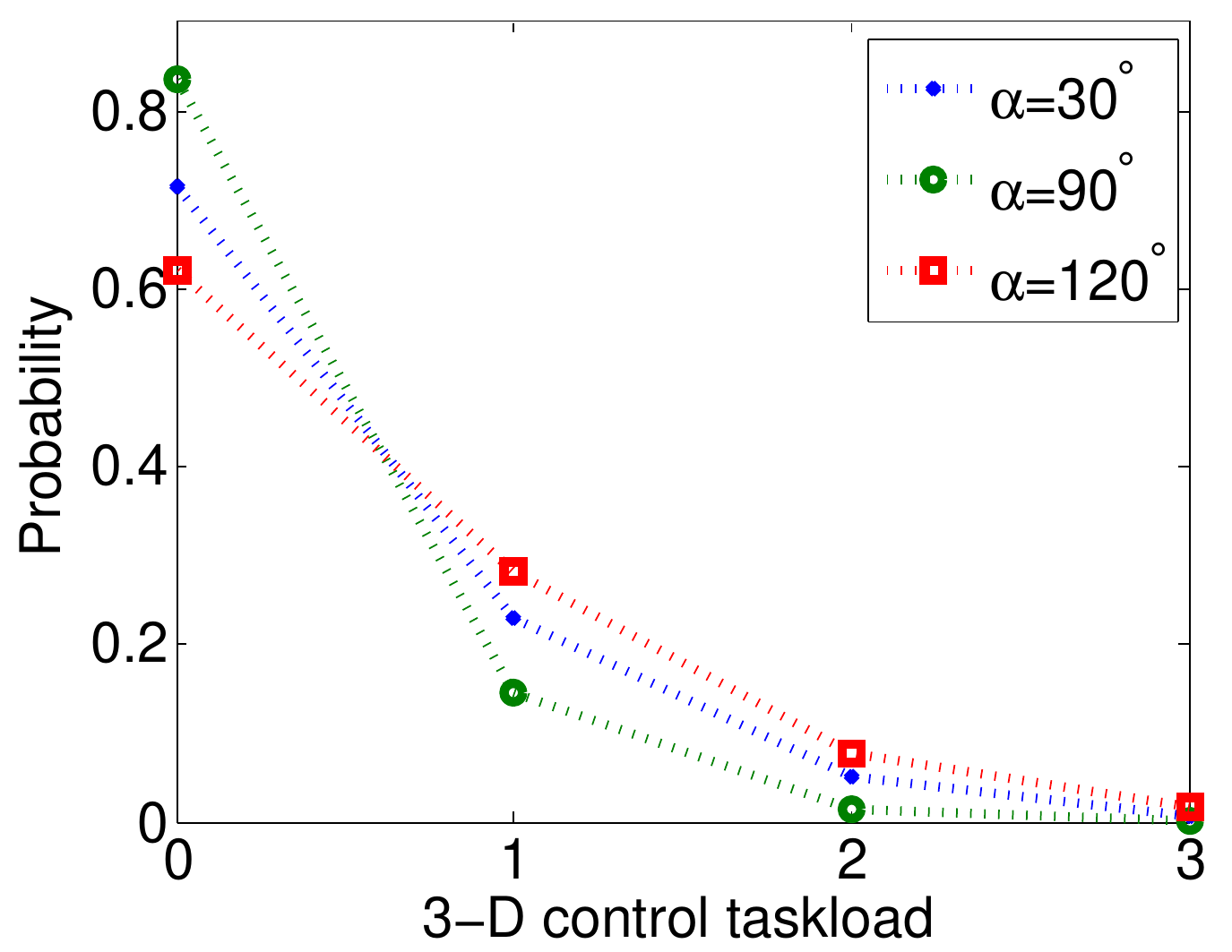}
}
\subfigure[Total taskload probability]{
\includegraphics[width=1.6in]{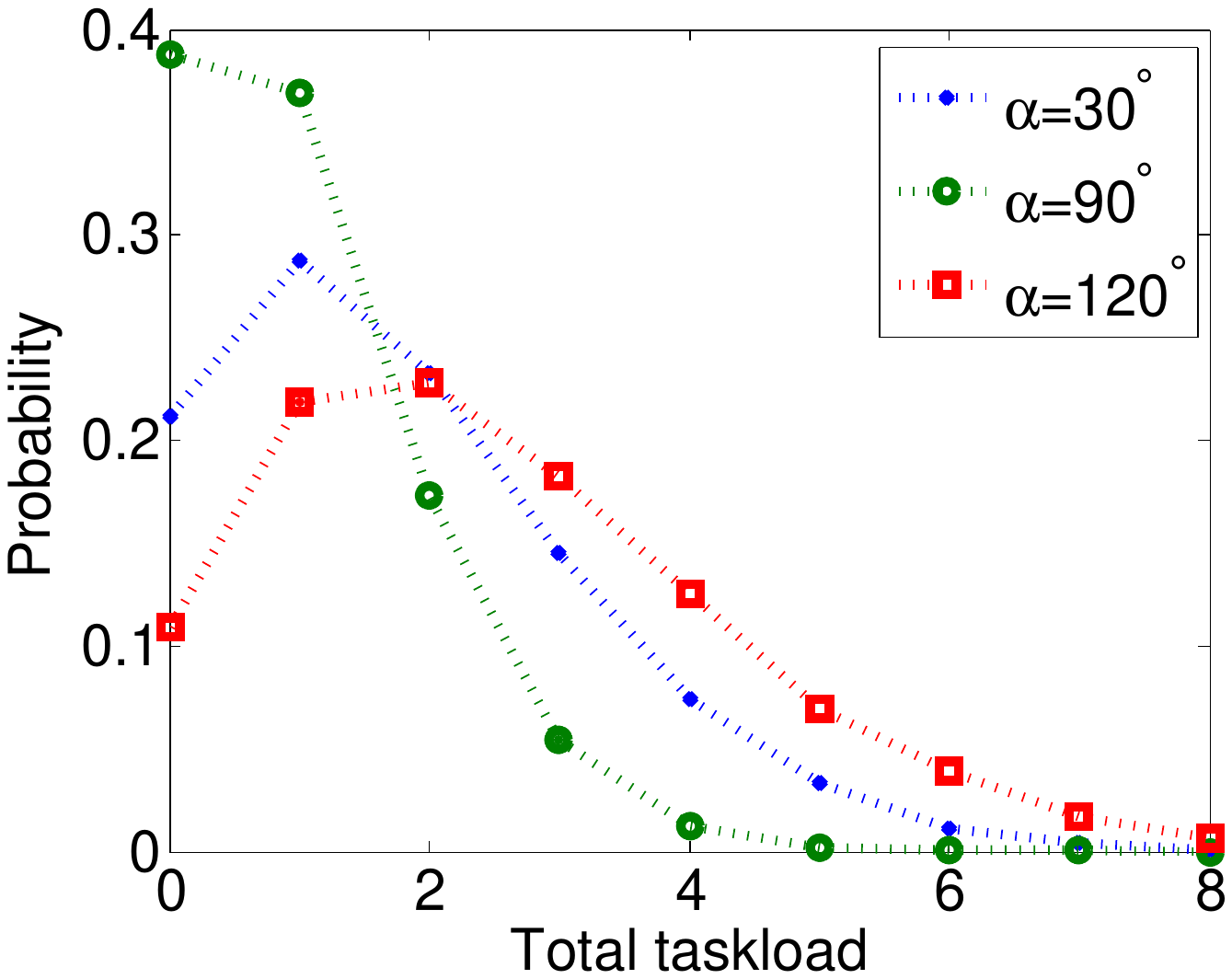}
}
\caption{Effect of crossing angle on 2h taskload probability; \emph{stringent} and \emph{severe} standard flows intersect}
\label{fig:stringent_severe_cross}
\end{figure}

Figure \ref{fig:stringent_severe_cross} shows the angle effect on a crossing between a \emph{stringent} and a \emph{severe} flow. When compared to Figure \ref{fig:stringent_cross}, the taskload probability decreases slightly (best seen for the $\alpha=90^{\circ}$ crossing)  because of the higher control tolerance in the second flow.

\begin{figure}[!h]
\centering
\subfigure[Deviation control taskload probability]{
\includegraphics[width=1.6in]{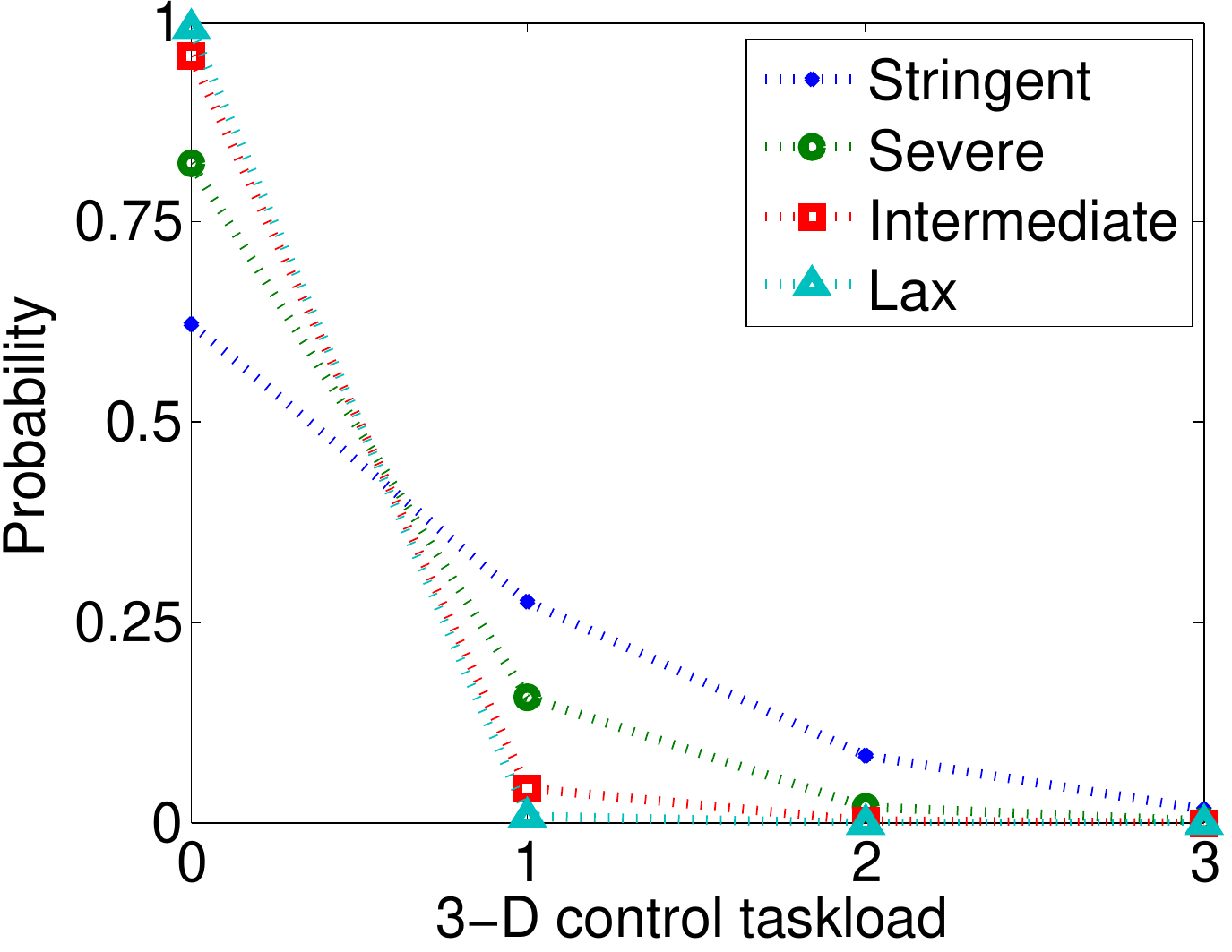}
}
\subfigure[Total taskload probability]{
\includegraphics[width=1.6in]{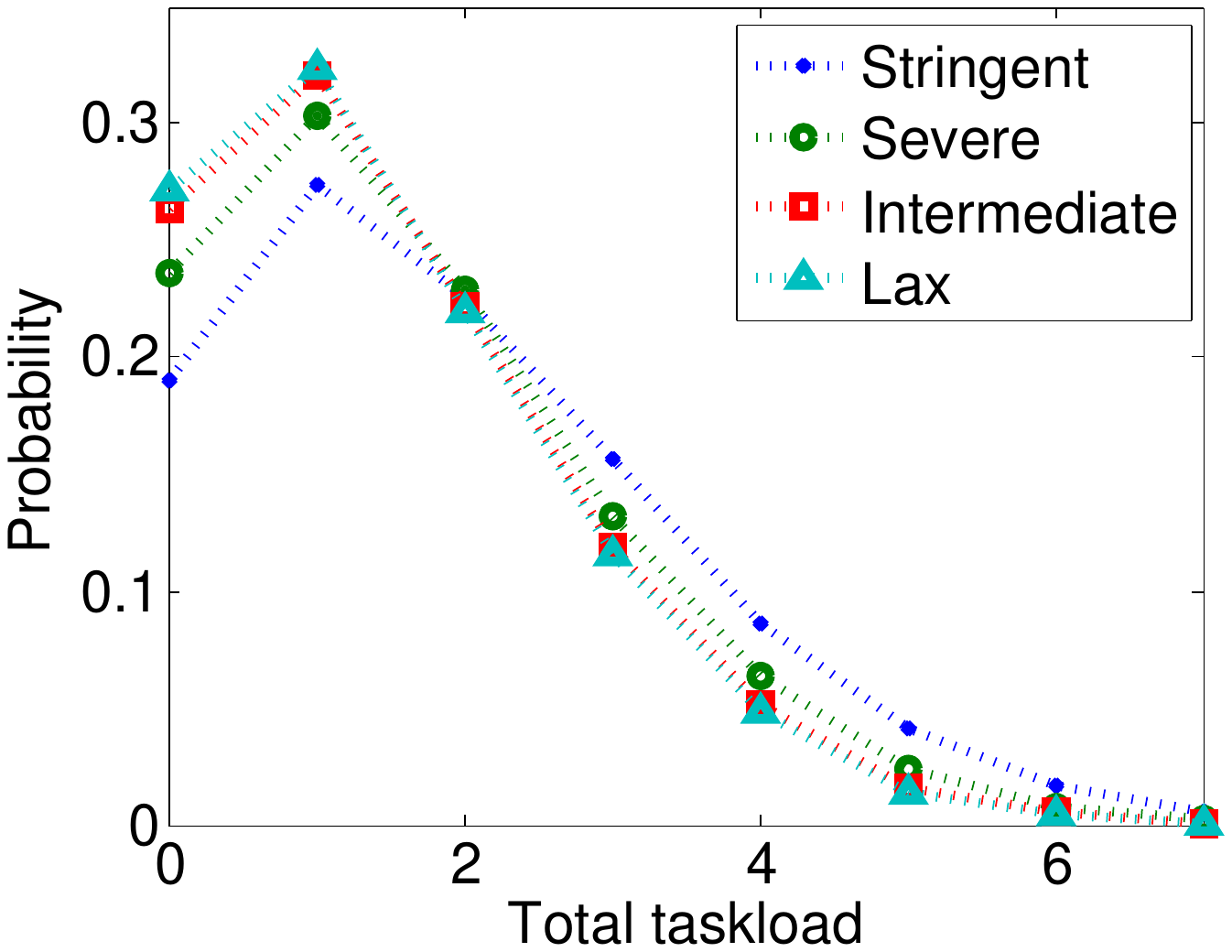}
}
\caption{Effect of deviation tolerance on 2h taskload probability; identical flows intersect at $\alpha=30^{\circ}$}
\label{fig:30_cross}
\end{figure}

\begin{figure}[!h]
\centering
\subfigure[Deviation control taskload probability]{
\includegraphics[width=1.6in]{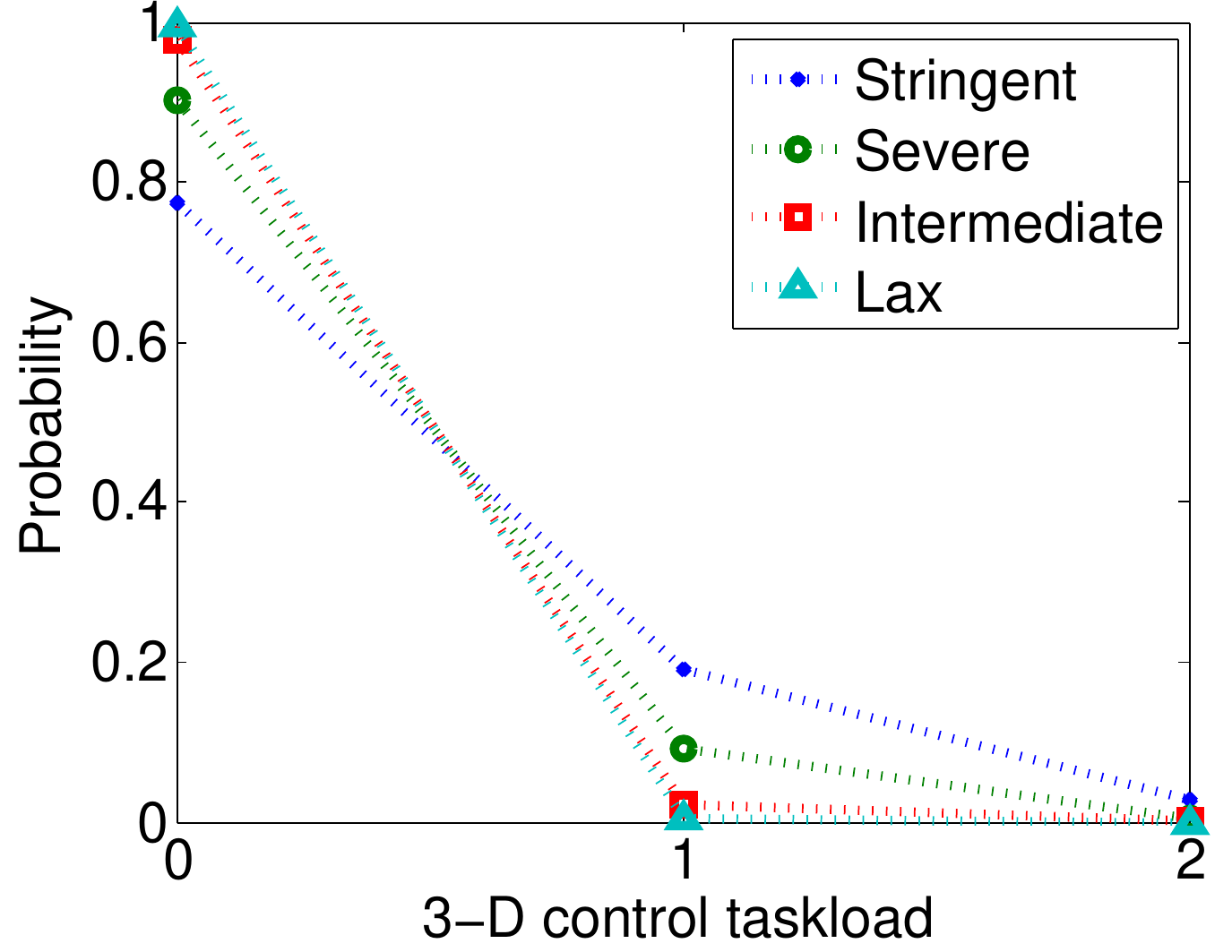}
}
\subfigure[Total taskload probability]{
\includegraphics[width=1.6in]{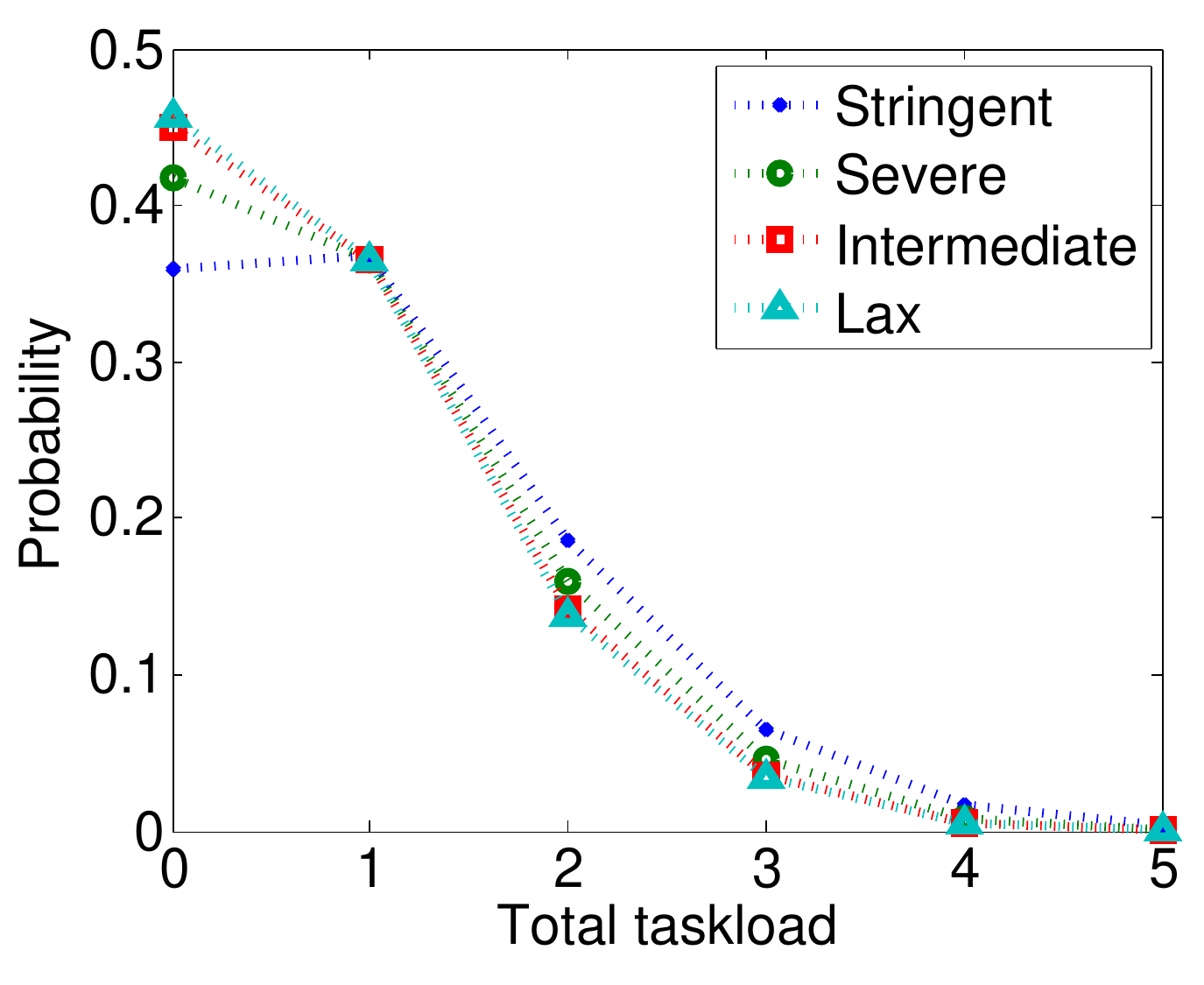}
}
\caption{Effect of deviation tolerance on 2h taskload probability; identical flows intersect at $\alpha=90^{\circ}$}
\label{fig:90_cross}
\end{figure}

\begin{figure}[!h]
\centering
\subfigure[Deviation control taskload probability]{
\includegraphics[width=1.6in]{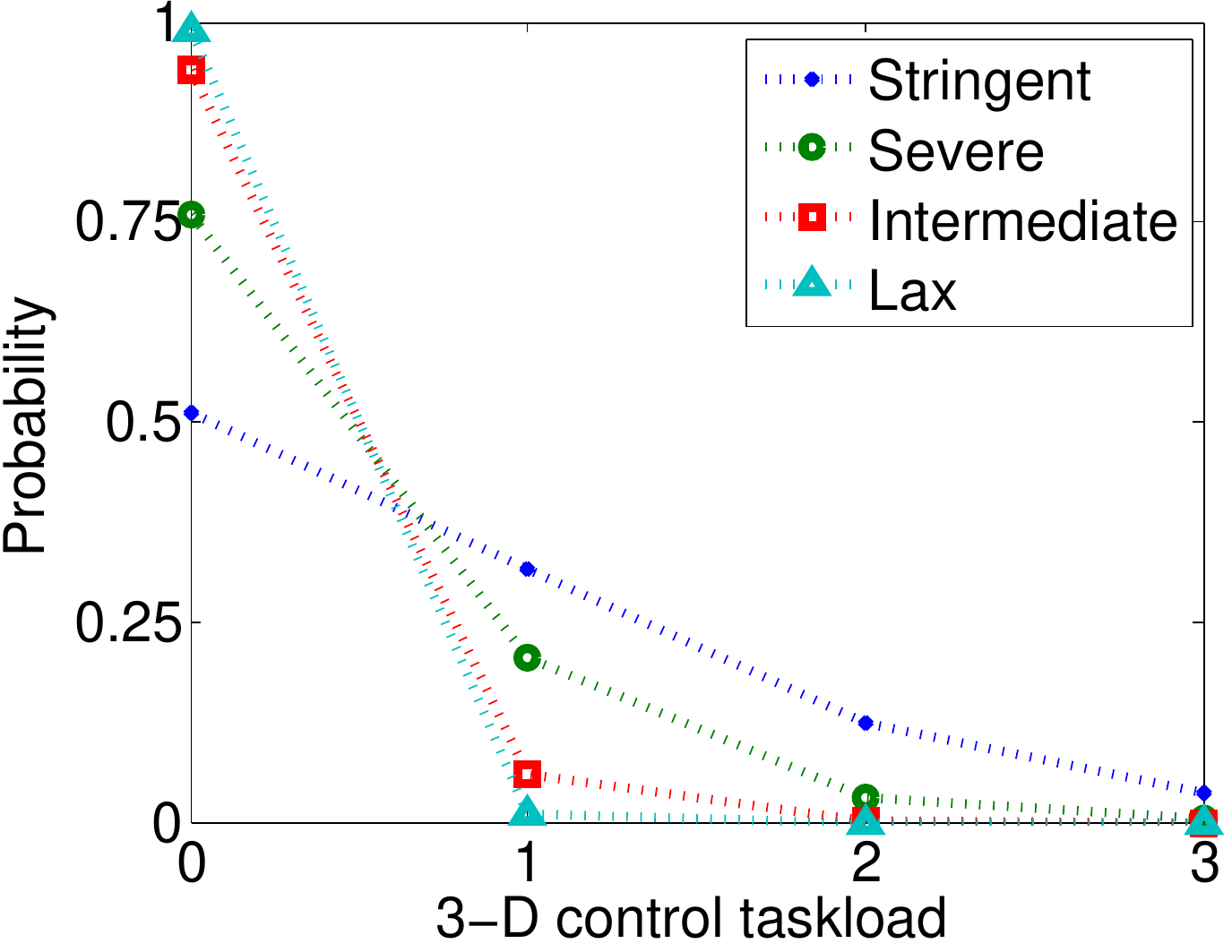}
}
\subfigure[Total taskload probability]{
\includegraphics[width=1.6in]{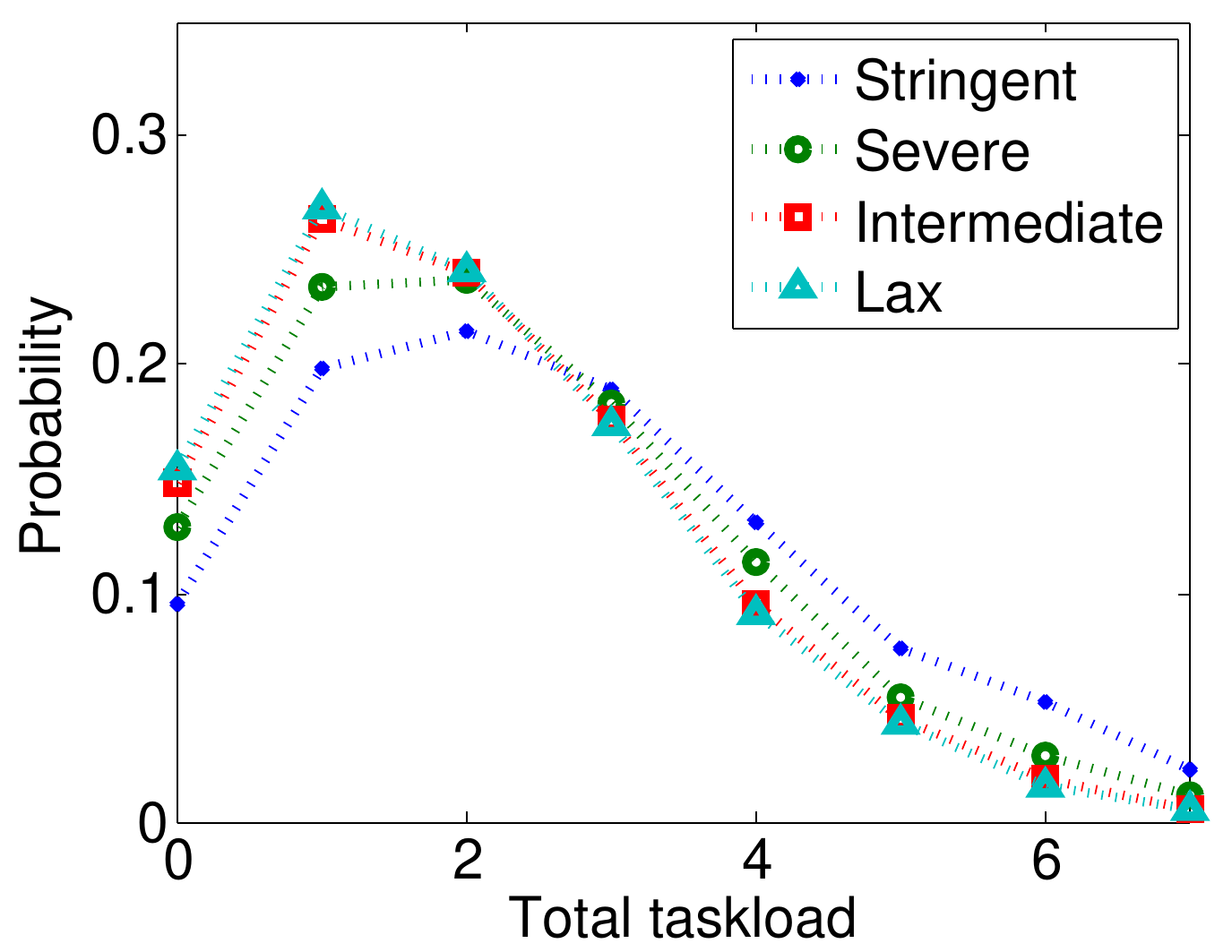}
}
\caption{Effect of deviation tolerance on 2h taskload probability; identical flows intersect at $\alpha=120^{\circ}$}
\label{fig:120_cross}
\end{figure}

Figures \ref{fig:30_cross}, \ref{fig:90_cross}, and \ref{fig:120_cross} show the effect of the control tolerance on the taskload at a crossing of $\alpha=30^{\circ}$, $\alpha=90^{\circ}$, and $\alpha=120^{\circ}$ respectively. As can be expected, for all the crossing values the control tolerance bounds play a significant role in the taskload levels. This effect is however least noted in the $\alpha=90^{\circ}$ crossing where the aircraft spend the least amount of time in the safe-zone. For the \emph{intermediate} and \emph{lax} tolerances, the taskload only comes from the conflict resolution interventions (the probability of no control interventions is close to 1). Since the control tolerance plays no part in the flow scheduling, the conflict resolution taskload has not been shown in plots other than in Figure \ref{fig:stringent_cross} 

\section{Conclusion}

In a context of NextGen flow corridors, this research seeks to quantify what controller taskload is required to maintain aircraft adherence to their 4-D trajectories, within certain tolerance levels.

Fundamental stochastic models of the aircraft motion (an Ornstein-Uhlenbeck process) and of the flow scheduling (a Poisson process) have been introduced, along with reasonable numerical values of the model parameters. The Ornstein-Uhlenbeck aircraft model was calibrated - by a least squares linear regression and by a maximum likelihood estimation - to match simulated data, obtained from a random number generator defined by a Johnson unbounded system $S_U$. The use of simulated data to calibrate the model was imposed by the absence of experimental Flight Technical Error (FTE) data; the random number generator provided fictitious data consistent with statistical studies of the aircraft motion. The Poisson flow model used intensity parameters given in recently published research. Analytic expressions have been derived for the taskload probability density functions for basic functional elements of the flow structure (a single lane, multiple parallel lanes, and flow corridor crossings). Monte Carlo simulations have been performed for these basic functional elements of the flow structure and the controller taskload probabilities have been exhibited.

It has been found that the current aircraft performance \emph{de facto} supports the flow corridor concept, by providing high-precision navigation. The controller taskload needed - assuming no anticipated reaction from the pilot - to prevent the aircraft from excessive deviation outside of their nominal 4-D trajectories was moderate, not exceeding ten interventions over two hours for a single or multiple lane flow with very high density traffic (a mean of 60 aircraft per hour). The crossing of flows however poses different concerns, where solving the occurring scheduling conflicts at the crossing point becomes the foremost source of controller taskload.

The authors acknowledge that experimental FTE data measurements and a more accurate statistical flow description are needed to improve the model accuracy and, in turn, to increase confidence in its predictions. Nevertheless, a coherent deductive taskload model that only requires simple fundamental microscopic (aircraft motion) and macroscopic (flow) models has been demonstrated in this research.

In future work, we plan on using the detailed Ornstein-Uhlenbeck model of the aircraft developed here to quantify risk severity in the case of extreme deviations from nominal operations. Studies of the risk propagation across flow geometry under degraded operations, and its dependency with the route structure will also be conducted.


\section*{Acknowledgements}

This work was supported by NASA under grant NNX08AY52A.

The authors thank Maxime Gariel from the Massachusetts Institute of Technology and Adan Vela, Erwan Sala\"{u}n from the Georgia Institute of Technology for their valuable contribution to this paper.





%

\bibliographystyle{IEEEtran}

\bibliography{taskload}

\begin{thebibliography}{10}
\providecommand{\url}[1]{#1}
\csname url@samestyle\endcsname
\providecommand{\newblock}{\relax}
\providecommand{\bibinfo}[2]{#2}
\providecommand{\BIBentrySTDinterwordspacing}{\spaceskip=0pt\relax}
\providecommand{\BIBentryALTinterwordstretchfactor}{4}
\providecommand{\BIBentryALTinterwordspacing}{\spaceskip=\fontdimen2\font plus
\BIBentryALTinterwordstretchfactor\fontdimen3\font minus
  \fontdimen4\font\relax}
\providecommand{\BIBforeignlanguage}[2]{{%
\expandafter\ifx\csname l@#1\endcsname\relax
\typeout{** WARNING: IEEEtran.bst: No hyphenation pattern has been}%
\typeout{** loaded for the language `#1'. Using the pattern for}%
\typeout{** the default language instead.}%
\else
\language=\csname l@#1\endcsname
\fi
#2}}
\providecommand{\BIBdecl}{\relax}
\BIBdecl

\bibitem{JPDO:conops07}
{Joint Planning and Development Office}, ``Concept of operations for the {Next
  Generation Air Transportation System},'' Tech. Rep. Version 2.0, June 2007.

\bibitem{PBN08}
\emph{Performance-based Navigation {(PBN)} Manual}, Third edition~ed.,
  International Civil Aviation Organization, 2008.

\bibitem{Radio08}
\emph{2008 Federal Radionavigation Plan}, {Department of Defense, Department of
  Homeland Security, and Department of Transportation} Std.

\bibitem{Mun07}
A.~Mundra and E.~Simons, ``Self-separation corridors,'' in \emph{26th Digital
  Avionics Systems Conference}, 2007.

\bibitem{You10}
A.~Yousefi, J.~Lard, and J.~Timmerman, ``{NextGen} flow corridors initial
  design, procedures, and display functionalities,'' in \emph{29th Digital
  Avionics Systems Conference}, 2010.

\bibitem{ICAO98}
\emph{Annex 11 – Air Traffic Services}, ICAO Std. Green pages, attachment B,
  paragraph 3.2.1., Rev. 12th edition incorporating amendments 1-38, July 1998.

\bibitem{Rei64}
P.~Reich, ``A theory of safe separation standards for air traffic control,
  technical report 64041,'' Royal Aircraft Establishment, UK, Tech. Rep., 1964.

\bibitem{Blo06}
H.~Blom, G.~Bakker, B.~K. Obbink, and M.~Klompstra, ``Free flight safety risk
  modelling and simulation,'' in \emph{Second International Conference on
  Research in Air Transportation}, 2006.

\bibitem{Irv01}
R.~Irvine, ``A geometrical approach to conflict probability estimation,'' in
  \emph{Fourth USA / Europe Air Traffic Management Seminar}, 2001.

\bibitem{PaielliErzberger:1997}
R.~Paielli and H.~Erzberger, ``Conflict probability estimation for free
  flight,'' \emph{Journal of Guidance, Control, and Dynamics}, vol.~20, no.~3,
  pp. 588--596, 1997.

\bibitem{Blin}
K.~Blin, M.~Akian, F.~Bonnans, E.~Hoffman, and K.~Zeghal, ``A stochastic
  conflict detection method integrating planned heading and velocity changes,''
  in \emph{IEEE Conference on Decision Control}, 2000.

\bibitem{Pra08}
M.~Prandini and J.~Hu, ``Application of reachability analysis for stochastic
  hybrid systems to aircraft conflict prediction,'' in \emph{47th IEEE
  Conference on Decision and Control}, 2008.

\bibitem{moreau2005}
D.~Moreau and S.~Roy, ``A stochastic characterization of en route traffic flow
  management strategies,'' in \emph{AIAA Guidance, Navigation, and Control
  Conference}, 2005, pp. 6274--6285.

\bibitem{Wan09}
Y.~Wan and S.~Roy, ``A scalable methodology for evaluating and designing
  coordinated air-traffic flow management strategies under uncertainty,''
  \emph{IEEE Transactions in Intelligent Transportation Systems}, vol.~9,
  no.~4, pp. 644--656, December 2009.

\bibitem{Sch77}
D.~K. Schmidt, ``On the conflict frequency at air route intersections,''
  \emph{Transportation Research}, vol.~11, no.~5, pp. 351--355, October 1977.

\bibitem{Sal10}
E.~Sala\"{u}n, M.~Gariel, A.~E. Vela, E.~Feron, and J.-P.~B. Clarke, ``Airspace
  complexity estimations based on data-driven flow modeling,'' in \emph{AIAA
  Guidance, Navigation and Control Conference}, 2010.

\bibitem{Dun75}
W.~J. Dunlay, ``Analytical models of perceived air traffic control conflicts,''
  \emph{Transportation Science}, vol.~9, no.~2, pp. 149--164, May 1975.

\bibitem{Jed08}
B.~G. Jeddi, ``A statistical separation standard and risk-throughput modeling
  of the aircraft landing process,'' Ph.D. dissertation, George Mason
  University, 2008.

\bibitem{Vel10}
A.~Vela, E.~Sala\"{u}n, M.~Gariel, E.~Feron, J.~Clarke, and W.~Singhose,
  ``Determining bounds on controller workload rates at an intersection,'' in
  \emph{American Control Conference}, 2010.

\bibitem{Leb00}
B.~Leblanc, O.~Renault, and O.~Scaillet, ``A correction note on the first
  passage time of an {Ornstein-Uhlenbeck} process to a boundary,''
  \emph{Finance and Stochastics}, vol.~4, no.~1, pp. 109--111, 2000.

\bibitem{Met10}
A.~Metzler, ``On the first passage problem for correlated {Brownian} motion,''
  \emph{Statistics and Probability Letters}, vol.~80, no. 5-6, pp. 277--284,
  March 2010.

\bibitem{Hun93}
M.~S. Huntley, J.~W. Turner, and R.~Palmer, ``Flight technical error for
  {Category B} non-precision approaches and missed approaches using
  non-differential {GPS} for course guidance,'' U.S. Department of
  Transportation, Volpe National Transportation Systems Center, Tech. Rep.,
  November 1993.

\bibitem{Wil05}
D.~M. Williams, M.~C. Consiglio, J.~L. Murdoch, and C.~H. Adams, ``Flight
  technical error analysis of the {SATS} higher volume operations simulation
  and flight experiments,'' in \emph{24th Digital Avionics Systems Conference},
  2005.

\bibitem{Lev03}
B.~S. Levy, P.~Som, and R.~Greenhaw, ``Analysis of flight technical error on
  straight, final approach segments,'' MITRE Corp., Tech. Rep., 2003.

\bibitem{Jon49}
N.~L. Johnson, ``Systems of frequency curves generated by methods of
  translation,'' \emph{Biometrika}, vol.~36, pp. 149--76, 1949.

\bibitem{Hil76}
I.~D. Hill, R.~Hill, and R.~L. Holder, ``Algorithm {AS 99}: Fitting {Johnson}
  curves by moments,'' \emph{Journal of the Royal Statistical Society. Series C
  (Applied Statistics)}, vol.~25, no.~2, pp. 180--189, 1976.

\bibitem{Win78}
K.~B. Winterbon, ``Determining parameters of the {Johnson SU} distribution,''
  \emph{Communications in Statistics - Simulation and Computation}, vol.~7,
  no.~3, pp. 223--226, 1978.

\bibitem{Sal11}
E.~Sala\"{u}n, M.~Gariel, A.~E. Vela, and E.~Feron, ``Aircraft proximity maps
  based on data-driven flow modeling,'' \emph{Submitted to the Journal of
  Guidance Control and Dynamics}, 2011.

\end{thebibliography}

\section*{Keywords}
\scriptsize{4-D trajectory, flow geometry, conflict avoidance, trajectory correction, Required Navigation Performance (RNP), Flight Technical Error (FTE), controller taskload, Ornstein-Uhlenbeck mean-reverting process, Wiener process, Brownian motion, Poisson process}

%

\section*{Author biographies}
\IEEEoverridecommandlockouts
\vspace*{-4\baselineskip}
\begin{IEEEbiographynophoto}{Vlad Popescu}
\scriptsize{received his Dipl\^{o}me d'Ing\'{e}nieur (2009) with a concentration in Applied Mathematics from the \'{E}cole Polytechnique, France, after attending the Lyc\'{e}e Louis le Grand in Paris (2004-2006). He is currently pursuing a PhD in Aerospace Engineering at the Georgia Institute of Technology in Atlanta, GA. 

His professional experiences include Onera in Meudon, France (2009) and the SESAR Joint Undertaking in Brussels, Belgium (2010). He is interested in applying control, dynamical systems, and game theory to cognitive engineering in air transportation.

He was a recipient of the Excellence-Major scholarship from the French Ministry of Foreign Affairs.}
\end{IEEEbiographynophoto}
\vspace*{-4\baselineskip}
\begin{IEEEbiographynophoto}{John-Paul B. Clarke}
\scriptsize{earned the S.B., S.M. and Sc.D. degrees in Aerospace Engineering from the Massachusetts Institute of Technology, Cambridge, MA in 1991, 1992, and 1997, respectively.

He is currently an Associate Professor in the Daniel Guggenheim School of Aerospace Engineering at the Georgia Institute of Technology, with a courtesy appointment in the H. Milton Stewart School of Industrial and Systems Engineering, and was previously a faculty member in the Department of Aeronautics and Astronautics at MIT.

His research and teaching address issues of optimization and robustness in aircraft and airline operations, air traffic management and the environmental impact of aviation. His work has garnered several awards including the AIAA/AAAE/ACC Jay Hollingsworth Speas Airport Award in 1999, the FAA Excellence in Aviation Award in 2003, and the National Academy of Engineering (NAE) Gilbreth Lectureship in 2006.

Dr. Clarke is a member of the AGIFORS, AIAA, INFORMS, ION and Sigma Xi.}
\end{IEEEbiographynophoto}
\vspace*{-4\baselineskip}
\begin{IEEEbiographynophoto}{Karen M. Feigh}
\scriptsize{holds a B.S. in Aerospace Engineering and a Ph.D. in Industrial and Systems Engineering from the Georgia Institute of Technology, USA, and a MPhil in Aeronautics from Cranfield University, UK.

She is currently an Assistant Professor in the Daniel Guggenheim School of Aerospace Engineering at Georgia Tech. She has previously worked on fast-time air traffic simulation, conducted ethnographic studies of airline and fractional ownership operation control centers, and designed expert systems for air traffic control towers.}
\end{IEEEbiographynophoto}
\vspace*{-4\baselineskip}
\begin{IEEEbiographynophoto}{Eric Feron}
\scriptsize{holds his B.S. (1989) from the \'{E}cole Polytechnique, his M.S. (1990) in Computer Science from the \'{E}cole Normale Sup\'{e}rieure, France, and his Ph.D. (1994) in Aerospace Engineering from Stanford University.

He has been the Dutton-Ducoffe Professor of Aerospace Software Engineering at the Georgia Institute of Technology since 2005, and was previously a faculty member in the department of Aeronautics and Astronautics at MIT from 1993 to 2005.

He is interested in using control, optimization and computer science for problems such as multi-agent operations for air traffic control systems and aerospace software certification.

He has published two books and is an advisor to the French Academy of Technologies.} 
\end{IEEEbiographynophoto}

\end{document}